\documentclass[onecolumn,floatfix,superscriptaddress,a4paper,showpacs,showkeys,nofootinbib,notitlepage]{revtex4-1}
\usepackage[colorlinks=true,linktocpage=true,linkcolor=blue,citecolor=blue,allcolors=blue]{hyperref}
\usepackage{epsfig}
\usepackage{latexsym}
\usepackage[utf8]{inputenc}
\usepackage{xspace}
\usepackage{indentfirst}
\usepackage{enumerate}
\usepackage{color}

\newenvironment{referee}{\par\color{blue}}{\par}

\graphicspath{{newfigs/},{NA61/},{figs/},{WIP/}}
\usepackage{amsmath}
\usepackage{amssymb}
\usepackage[english]{babel}
\usepackage{url}

\newcommand{\mean}[1]{\langle #1 \rangle}

\newcommand{\Eq}[1]{Eq.~(\ref{#1})}

\newcommand\ddfrac[2]{\frac{\displaystyle #1}{\displaystyle #2}}

\newcommand{\rom}[1]{\uppercase\expandafter{\romannumeral #1\relax}}

\newcommand{\eq}[1]{\begin{align} #1 \end{align}}

\newcommand{\kurt}[0]{\kappa\sigma^2}

\begin{document}

 \title{Binomial acceptance corrections\\ for particle number distributions in high-energy reactions}

\author{Oleh Savchuk}
\affiliation{Taras Shevchenko National University of Kyiv, 03022 Kyiv, Ukraine}

\author{Roman V. Poberezhnyuk}
\affiliation{Bogolyubov Institute for Theoretical Physics, 03680 Kyiv, Ukraine}
\affiliation{Frankfurt Institute for Advanced Studies, Giersch Science Center,
D-60438 Frankfurt am Main, Germany}
\author{Volodymyr~Vovchenko}
\affiliation{Institut f\"ur Theoretische Physik,
Goethe Universit\"at Frankfurt, D-60438 Frankfurt am Main, Germany}
\affiliation{Frankfurt Institute for Advanced Studies, Giersch Science Center,
D-60438 Frankfurt am Main, Germany}
\author{Mark I. Gorenstein}
\affiliation{Bogolyubov Institute for Theoretical Physics, 03680 Kyiv, Ukraine}
\affiliation{Frankfurt Institute for Advanced Studies, Giersch Science Center, D-60438 Frankfurt am Main, Germany}

\date{\today}

\begin{abstract}
The binomial acceptance correction procedure is studied for particle number distributions  detected in high energy reactions in finite regions of the momentum space. 
We present acceptance correction formulas 
for scaled variance, skewness, and kurtosis.  
Our considerations include various specific types of  particles -- positively or negatively charged, baryons and antibaryons
--  as well as conserved charges, namely, the net baryon number and electric charge.
A simple model with effects of exact charge conservation, namely the Bessel distribution, is studied in some detail where effects of multi-particle correlations are present. 
The accuracy of the binomial filter is studied with UrQMD model simulations of inelastic proton-proton reactions. 
Binomial acceptance correction procedure works well when used inside a small region of phase space as well as for certain other types of corrections, in particular for constructing net proton fluctuations from net baryon ones. 
Its performance is less accurate when applied to obtain UrQMD  fluctuations in a finite rapidity window from fluctuations in the full $4\pi$ space.

\end{abstract}


\maketitle

\section{Introduction}

Investigation of the  phase diagram of strongly interacting matter is today
one of the most important topics in nuclear and particle physics.
Transitions between different phases of this matter
are expected to reveal themselves as specific patterns in particle number fluctuations. 
In particular, a critical point (CP) should yield large deviations of the conserved charges from their respective baselines in finite regions of the phase space around a CP, showing universal signals in various high-order susceptibilities~\cite{Stephanov:1998dy,Stephanov:1999zu,Athanasiou:2010kw,Stephanov:2008qz,Kitazawa:2012at,Vovchenko:2015uda}.
This generally applies not only to the hypothetical chiral QCD CP which has garnered most attention, but also to the better established CP of the nuclear liquid-gas transition ~\cite{Sauer:1976zzf,Pochodzalla:1995xy}, which entails characteristic patterns in nucleon number fluctuations~\cite{Vovchenko:2015pya} as well as nuclear fragment distributions~\cite{Bondorf:1995ua}.

The particle number fluctuations can be characterized by the central moments, $\langle(\Delta N)^{2}\rangle\equiv \sigma^2$, $\langle (\Delta N)^{3}\rangle$, $\langle (\Delta N)^{4}\rangle$, etc, where $\langle...\rangle$ denotes the event-by-event averaging and $\Delta N \equiv N -\langle N \rangle$. The scaled variance $\omega$, as well as (normalized) skewness $S\sigma$ and kurtosis $\kappa\sigma^{2}$ of particle number distribution are defined as the following combinations of the central moments,
\eq{ \omega[N] & =\frac{\sigma^2}{\mean{N}}~,\label{omega}\\
S\sigma[N] & = \frac{\langle(\Delta N)^{3}\rangle}{\sigma^{2}}~, \label{skew}\\
 \kappa  \sigma^{2}[N] & = \frac{\langle(\Delta N)^{4}\rangle-3\langle(\Delta N)^{2}\rangle^2}{\sigma^{2}}~.\label{kurt}
 }
These can also be expressed through the cumulants $\kappa_n$ of the $N$-distribution: 
$\mean{N}=\kappa_1,~~ \omega =\kappa_2/\kappa_1,~~  S\sigma=\kappa_3/\kappa_2,~~ \kappa\sigma^{2}=\kappa_4/\kappa_2.$
The quantities (\ref{omega})-(\ref{kurt}) are the well known size-independent (intensive) measures of particle number fluctuations.

Besides the particle number fluctuations, the susceptibilities of conserved charges such as net baryon number $B$ and electric charge $Q$ are of  special interest. 
In thermodynamic equilibrium, they are connected to the grand canonical partition function
and thus contain information about the QCD equation of state. 
Namely, the cumulants of conserved charge distributions are calculated as the corresponding derivatives of the system pressure $p$:
\eq{\label{pres}
\kappa_n[Q_i]=VT^3~\frac{\partial^n (p/T^4)}{\partial (\mu_{Q_i}/T^4)^n}~,
}
where $V$ and $T$ are  the system's volume and temperature,  $Q_i=B,~Q$, and $\mu_B$, $\mu_Q$ are, respectively, the baryon and electric chemical potentials.
Having generally longer equilibration times~\cite{Asakawa:2000wh,Jeon:2000wg}, the fluctuations of conserved charges are also thought to reflect properties of earlier stages of collision~\cite{Kitazawa:2012at}.
Studies of the higher-order fluctuation measures are motivated by their larger sensitivity to critical phenomena. Cumulants of a higher order are proportional to increasing powers of the correlation length $\xi$, and they are considerably more sensitive probes to the proximity of the CP than the variance~\cite{Stephanov:2008qz,Stephanov:2011pb}, as has been illustrated in a number of model calculations for net baryon and/or net charge fluctuations~\cite{Schaefer:2011ex,Vovchenko:2015pya,Chen:2015dra,Poberezhnyuk:2019pxs}.
Experimental studies of such fluctuation measures are in progress~\cite{Luo:2017faz}.
This motivates our study of acceptance effects for the higher-order fluctuation measures.

Of course, 
the baryon number and electric charge are globally conserved in high energy collisions, meaning that these quantities do not fluctuate in the full phase space provided that the events under consideration have the same number of participants.
Baryon number and electric charge of the entire system are conserved event-by-event.
Therefore, actual fluctuations of conserved charges can only be seen by considering finite acceptance regions. 
The optimal choice of acceptance for comparing the measurements with predictions of equilibrium thermodynamics in the grand-canonical ensemble is not trivial.
If acceptance is too small, the trivial Poisson-like
fluctuations dominate~\cite{Athanasiou:2010kw}.
The acceptance should be large enough compared to correlation lengths relevant for various physics processes, in particular, those related to the QCD CP~\cite{Ling:2015yau}.

A crucial question is connecting the quantities (\ref{omega}), (\ref{skew}), and (\ref{kurt}) measured in finite regions of the momentum space with predictions of various physical models.
In the present paper
acceptance effects are modeled by the binomial distribution. 
Namely, the binomial acceptance corrections~(BAC) assume that each particle of the $i$th type is accepted by detector with a fixed probability $x_i$. This probability $0\leq x_i=\langle n_i\rangle/\langle N_i\rangle \leq 1$ equals the ratio of the mean value $\langle n_i\rangle $
of accepted particles to that of $\langle N_i \rangle $  of all particles of the $i$-th type. 
The main assumption of the binomial acceptance is that the probability $x_i$ is the same for all particles of a given type and independent of any properties of a specific event. 
This assumption allows to relate the cumulants
within a finite acceptance to their values in the larger, encompassing phase space. 
We will use the method of characteristic functions which was used previously 
for similar purposes in Ref.~\cite{Bzdak:2012ab}. 
The present formalism does recreate the prior results on the BAC~\cite{Bzdak:2012ab,Kitazawa:2011wh,Kitazawa:2016awu}, as one would expect.
Our main focus here is on a number of special cases for which the present formalism is found to be most suitable.
We will consider both, the fluctuations of the specific particle species and also that of globally conserved charges such as baryon number or electric charge. 
We analyze the performance of the binomial filter for acceptance corrections in the momentum space as well for constructing net proton fluctuations from net baryon ones.
For that we use ultra-relativistic quantum molecular dynamics~(UrQMD) model~\cite{Bass:1998ca,Bleicher:1999xi} simulations of inelastic p+p interactions.

The paper is organized as follows. In Sec.~\ref{sec-binomial} we present the formulas of the BAC
which connect the fluctuations measures
in the finite $x$-acceptances with the corresponding quantities  
in the full phase space. 
Sec.~\ref{sec-poisson} presents the typical multiplicity distributions in grand-canonical and canonical statistical mechanics of relativistic particles.
In Sec.~\ref{sec-urqmd} the BAC performance is confronted with the 
the UrQMD model simulations.
Summary in Sec.~\ref{summary} closes the article.

\begin{widetext}

\section{Binomial acceptance corrections}\label{sec-binomial}

Let the $P(N)$ function denote a normalized probability distribution for observing  $N$ particles of a given type in the full phase space. 
The BAC for particle number fluctuations  assume that the probability $p(n,x)$ to observe $n$ particles detected in the finite $x$-region of the phase space 
is given as
\eq{\label{binomial}
p(n,x)=   
\sum_{N=n}^{\infty} \frac{N!}{n!(N-n)!}\,x^n (1-x)^{N-n}\,P(N)~\equiv \sum_{N=n}^{\infty} B(N,n;x)\,P(N)~.
}

\subsection{BAC for the particle number fluctuation}\label{subsec-pn}
First, we consider the BAC (\ref{binomial}) applied to particles of a given type. The characteristic function of the $P(N)$ distribution is defined as
\eq{\label{FN1}
F_{N}(k) =\langle e^{ikN}\rangle
=
\sum_{N=0}^{\infty}e^{ikN}P(N)=\exp\left[\sum_{l=1}^{\infty}\kappa_{l}[N]\frac{(ik)^l}{l!} \right]~,
}
where $\kappa_{l}[N]$ is the $l$-th cumulant of the distribution $P(N)$. 
The corresponding characteristic function for the number of accepted particles reads
\eq{\nonumber
  f_{n}(k|x) &=   \sum_{n=0}^{\infty}e^{ikn}~ p(n|x)=\sum_{N=0}^{\infty}P(N)\sum_{n=0}^{n_0}e^{ikn}~B(N,n|x)\\ &=\sum_{N=0}^{\infty}P(N)(1-x+x~e^{ik})^{N}=\sum_{N=0}^{\infty}P(N)e^{\phi[k|x]{N}}
  =F_{N}(-i\phi[k|x])~,
  \label{fn}
}
where 
$\phi[k|x] \equiv {\rm ln}\left(1-x+x e^{ik}\right)$ is the cumulant generating function of binomial distribution
and $F_N$ is given by  Eq.~(\ref{FN1}). We checked that such procedure is correct for discrete random variables.

The acceptance parameter $0\leq x\leq 1$ has a simple meaning $x=\langle n\rangle/\langle N\rangle$, i.e., it equals to the ratio of the average multiplicities of the accepted and all particles.    
At $x \rightarrow  1$ one finds $f_n(k|x) \cong F_N(k|x)$, i.e., $p(n|x)\cong P(n)$.
\eq{\label{ch-poisson}
f_n(k|x) \stackrel{x \to 0}{\simeq} 1 - x\mean{N}(1-e^{ik}) \stackrel{x \to 0}{\simeq} \exp[x\mean{N}(e^{ik}-1)]~,
}
which is a characteristic function of the Poisson distribution with a mean equal to $x\mean{N}$.

The cumulants of the $p(n,x)$ probability distribution are calculated as
\eq{\label{kl}
\kappa_l[n|x] = \left(\frac{d}{d(ik)}\right)^l\ln[f_n(k|x)]{\Big|_{k=0}}~.
}
The scaled variance, skewness, and kurtosis for the distribution (\ref{binomial}) of the accepted particles 
are then presented as follows:
\eq{
\label{w-x}
  \omega_x[n]  & \equiv \frac{\kappa_2[n|x]}{\kappa_1[n|x]}= 1-x+x\omega[N]~,\\
\label{s_x}
 S\sigma_x[n] & = \frac{\kappa_3[n|x]}{\kappa_2[n|x]}=\frac{\omega[N]}{\omega_x[n]}\left\{ x^{2}S\sigma[N] + 3x(1-x) \right\} +\frac{1-x}{\omega_x[n]}(1-2x)~,\\
\label{k_x}
\kappa\sigma^{2}_x[n]  & = 
 \frac{\kappa_4[n|x]}{\kappa_2[n|x]}  =
    \ddfrac{\omega[N]}{\omega_x[n]}\left\{ x^{3}\kappa\sigma^{2}[N] + 6x^{2}(1-x)S\sigma[N]  +x(1-x)(7-11x) \right\}+\ddfrac{1-x}{\omega_x[n]}(1-6x(1-x)),
}
where
\eq{\label{N}
\omega[N]=\frac{\kappa_2[N]}{\kappa_1[N]}~,~~~~~ S\sigma[N]=\frac{\kappa_3[N]}{\kappa_2[N]}~,~~~~\kappa\sigma^2[N]=\frac{\kappa_4[N]}{\kappa_2[N]}~.
}
Equation (\ref{w-x}) was previously obtained in Refs.~\cite{Begun:2004gs,Braun-Munzinger:2016yjz}.
At $x\rightarrow 1$ in Eqs.~(\ref{w-x})-(\ref{k_x}), one evidently finds $\omega_x[n]\cong \omega[N]$, $S\sigma_x[n]\cong S\sigma[N]$, and $\kappa\sigma^2_x[n]\cong \kappa \sigma^2[N]$, i.e.,  the BAC fluctuation 
measures approach those in the full phase space.
In the opposite limit, $x \to 0$, the cumulant ratios are Poissonian, i.e. $\omega_x[n] \cong S\sigma_x[n] \cong \kappa\sigma^2_x[n]\cong 1$.

Equations (\ref{w-x})-(\ref{k_x}) can be reversed to express the fluctuations in full phase space in terms of fluctuations within a given binomial acceptance $x$:
\eq{\label{w_x[N]_reversed}
 \omega[N]& =1-\frac{1-\omega_x[n]}{x}~,\\
\label{s_x[N]_reversed}
 S\sigma[N] & =\frac{\omega[n]}{x^2\omega[N]}S\sigma[n] -\frac{1-x}{x^2\omega[N]}(1-2x+3x\omega[N])~,\\
\label{k_x[N]_reversed}
    \kappa\sigma^{2}[N] & = \ddfrac{\omega[n]}{x^3\omega[N]}\kappa\sigma^{2}[n] -\ddfrac{1-x}{x^3\omega[N]}\left\{1-6(1-x)x+(7-11x+6xS\sigma[N])x\omega[N]\right\}~.
}
It should be noted, however, that these ``inverse'' relations are to be used with care. 
By definition, both $\omega[N]$ and $\omega_x[n]$ are non-negative quantities. 
Equation (\ref{w-x}) guaranties
that $\omega_x[n]\ge 0$ at $0\le x\le 1$ for any non-negative value of $\omega[N]$. For the reverse relation (\ref{w_x[N]_reversed}), however, this is not guaranteed: 
The values of $0\le \omega_x[n]< 1-x$ are transformed by 
Eq.~(\ref{w_x[N]_reversed}) to a meaningless negative value of $\omega [N]$.
Similar arguments can be applied to higher order cumulants.

Let us consider two particular examples of the $P(N)$ distribution. A first example is a Poisson distribution,
\eq{\label{poiss}
P(N)= \exp(-\langle N\rangle)\,\frac{\langle N\rangle^N}{N!}~.
}
This distribution may correspond, e.g., to an equilibrium system of non-interacting Maxwell-Boltzmann particles in the grand canonical ensemble.
One  finds
\eq{\label{fl-poiss}
\omega[N]=S\sigma[N]=\kappa\sigma^2[N]=~1~
}
for the fluctuations in the full phase space whereas 
Eqs.~(\ref{w-x})-(\ref{k_x}) give
\eq{\label{x-poiss}
\omega_x[n]= S\sigma_x[n]= \kappa\sigma^{2}_x[n] =~ 1~
}
for fluctuations within acceptance.
The BAC quantities (\ref{x-poiss}) are independent of the $x$-acceptance parameter and equal to the fluctuation measures (\ref{fl-poiss}) in the full phase space. 
This last property of the BAC procedure is a unique feature of the Poisson distribution (\ref{poiss}).

As our second example we assume that the number of particles in the full phase space is fixed, i.e., 
\eq{\label{N0}
P(N)~=~\delta(N-N_0)~.
}
Such a scenario is approximately valid for the number of baryons in 
$p+p$ and nucleus-nucleus reactions at small and intermediate collision energies where the production of baryon-antibaryon pairs is negligible.

One finds, 
\eq{\label{N-fix}
\omega[N]=0~,~~~S\sigma[N]=-1~,~~~~\kappa\sigma^{2}[N]=1~.
}
 Equations~(\ref{w-x})-(\ref{k_x}) then correspond  to the  binomial probability distribution $B(n,N,x)$, giving
\eq{\label{bin}
\omega_x[n]=1-x~,~~~~ S\sigma_x[n]=~1-2x~,~~~~ \kappa\sigma^2_x[n]= 1-6x(1-x)~   .
}

\subsection{BAC for 
conserved charge fluctuations}\label{subsec-Q}
In this subsection, we consider the BAC for fluctuations of conserved charges.
We use notations $N_+$, $N_-$ and  $n_{+}$, $n_{-}$, for positively and negatively charged particles in the full space and in the $x$-acceptance region, respectively. 
Here conserved charge may correspond to any integer conserved number carried by hadrons, for instance the electric charge or baryon number.
Without loss of generality, we focus here on the net electric charge.
The nonzero values of electric charge and baryon number of final state hadrons detected in high energy collisions are $\pm 1$. 
Therefore, the net charge $Q$ is straightforwardly connected to the number of positively and negatively charged particles:
$Q=N_{+}-N_{-}=const$ in the full space and $q=n_+-n_-$ within the acceptance.

The distribution function of $N_+$ and $N_-$ can be presented in the following general form: 
\eq{\label{pN+N-}
{\cal P}(N_{+},N_{-})=\delta(N_{+}-N_{-}-Q)~P(N_{\rm ch}),
}
where $N_{\rm ch}\equiv N_{+}+N_{-}$.
The BAC are introduced as 
\eq{
p(n_+,n_-|x_{+},x_{-})=\sum_{N_{+},N_-=0}^{\infty}{\cal P}(N_+,N_-)B(N_+,n_+|x_+)B(N_-,n_-|x_-)~,
}
where the binomial distributions are defined in \Eq{binomial}. 
 The parameter  $x_+$ is defined as a ratio of the mean number of accepted (measured) positively charged particles to the mean number of all produced positively charged particles in a given sample of the collision events, $x_+=\langle n_+\rangle/\langle N_+\rangle$. Similarly, $x_-=\langle n_-\rangle/\langle N_-\rangle$. In the BAC procedure these quantities coincide with the probabilities for a randomly chosen positively (negatively) charged particle to end up within the detector acceptance. 
The characteristic function for the distribution of net charge $q = n_+ - n_-$ in the acceptance can be calculated as follows:
\eq{
f_q(k|x_+,x_-)& \equiv 
\sum_{n_{+},n_{-}=0}^{\infty}e^{ik(n_{+}-n_{-})}~p(n_+,n_-|x_{+},x_{-}) \nonumber\\
&=
(1-x_+ +x_+ e^{ik})^{Q/2}\,(1-x_-+x_- e^{-ik})^{-Q/2} ~F_{N_{\rm ch}}[-i\Phi(k|x_+,x_-)]~.
\label{FQ}
}
Here 
\eq{
\Phi(k|x_+,x_-) \equiv  
\frac{1}{2}~ \left[{\rm ln}(1-x_++x_+e^{ik})+{\rm ln}(1-x_-+x_-e^{-ik})\right]~
}
and 
$F_{N{\rm ch}}$ is the
characteristic function of the full 
space charged multiplicity distribution $P(N_{\rm ch})$
\eq{\label{F-charged}
F_{N_{ch}}(k) \equiv \sum_{N_{ch}=Q}^{\infty}P(N_{ch})e^{ikN_{ch}}=\exp\left[\sum_{l=1}^{\infty}\kappa_{l}[N_{ch}]\frac{(ik)^l}{l!} \right]~.
}

The $l$-th BAC cumulant of the net charge, $q$, fluctuations reads
\eq{\label{k_n_BAC}
\kappa_l[q|x_+,x_-] = \left(\frac{d}{d(ik)}\right)^l\ln[f_q(k|x_+,x_-)]{\Big|_{k=0}}~.
}
The leading four cumulants read
\eq{\label{k1x1x2}
\kappa_1[q] &= \mean{q}=x_+\mean{N_+}-x_-\mean{N_-},\\
\kappa_2[q] &= \xi_2^+\mean{N_+}+\xi_2^-\mean{N_-}+\left(\frac{\Delta x}{2}\right)^2\kappa_{2}[N_{ch}]\\\label{k3x1x2}
\kappa_3[q] &= \xi_3^+\mean{N_+}-\xi_3^-\mean{N_-} - \frac{3}{4}\Delta x\left[\xi_2^-+\xi_2^+\right]\kappa_{2}[N_{ch}] -\left(\frac{\Delta x}{2}\right)^3 \kappa_{3}[N_{ch}]\\
\kappa_4[q] &= \xi_4^+\mean{N_+} +   \xi_4^-\mean{N_-} + \frac{3}{4}\left[\xi_2^++\xi_2^-\right]^2 \kappa_{2}[N_{ch}]+  \Delta x\left[\xi_3^--\xi_3^+\right] \kappa_{2}[N_{ch}]+ \nonumber \\ 
& \quad + 3\left(\frac{\Delta x}{2}\right)^2\left[\xi_2^++\xi_2^-\right] \kappa_{3}[N_{ch}]+\left(\frac{\Delta x}{2}\right)^4 \kappa_{4}[N_{ch}]\label{k4x1x2}~.
}
Here $\mean{N_+}=(\mean{N_{ch}}+Q)/2$, $\mean{N_-}=(\mean{N_{ch}}-Q)/2$, ~$\Delta x = x_- - x_+$,
and

\eq{\label{cum_binomial}
\xi_1^\pm=x_\pm ,~~~~~~~~\xi_2^\pm =x_\pm (1-x_\pm),~~~~~~~~\xi_3^\pm=\xi_2^\pm (1-2x_\pm),~~~~~~~~\xi_4^\pm=\xi_2^\pm(1-6\xi_2^\pm)~.
}
 As seen from Eqs.~(\ref{k1x1x2})-(\ref{k4x1x2}), the BAC  {\it net charge} cumulants are calculated in terms of the cumulants 
 of the $P(N_{\rm ch})$ distribution of charged multiplicity in the full phase space.
In the case of equal acceptance parameters, $x_+=x_-\equiv x$, the cumulant ratios are simplified to
\eq{\label{w_x[Q]}
 \omega_x[q]
 & \equiv \frac{\kappa_2[q]}{\kappa_1[q]}= \frac{\mean{N_{ch}}}{Q}(1-x)~,\\
 \label{S[Q]}
S\sigma_x[q] & \equiv \frac{\kappa_3[q]}{\kappa_2[q]}= \frac{Q}{\mean{N_{ch}}}(1-2x)~,\\
    \kappa\sigma^{2}_x[q] & \equiv \frac{\kappa_4[q]}{\kappa_2[q]} =1+3x(1-x)\,( \omega[N_{ch}]-2)~.\label{kq}
}
We will use UrQMD simulations further on to analyze to what extent the assumption $x_+=x_-$ holds in realistic situations.

The above results can be straightforwardly generalized for the case of net baryon number fluctuations.
This is achieved through the following substitutions in Eqs.~(\ref{w_x[Q]}) and
(\ref{kq}): 
$q\rightarrow b$, $Q\rightarrow B$, $N_{\rm ch}\rightarrow N_B+N_{\overline{B}}$.
For sufficiently small collision energies in $p+p$ and nucleus-nucleus reactions one has $N_{\overline{B}}\ll N_B$, meaning that the number of baryons $N_B$ is approximately equal to the net baryon number $B$. 
Therefore, $(N_B+N_{\overline{B}})/B\cong 1$,  $\omega[N_B +N_{\overline{B}}]\cong \omega[N_B]\cong 0 $ 
and Eqs.~(\ref{w_x[Q]}),~(\ref{kq}) reduce to Eqs.~(\ref{bin}).

It should be noted that the scaled variance~(\ref{w_x[Q]}) and the skewness~(\ref{S[Q]}) exhibit a special behavior for the case of $e^++e^-$ and/or $p+\overline{p}$ reactions.
In these reactions all globally conserved charges are equal to zero, and thus $\omega_x[q]\equiv \infty$ and $S\sigma_x[q]\equiv 0$. 
On the other hand, the kurtosis (\ref{kq}) attains non-trivial values for all types of reactions.
\begin{figure*}[h]
\includegraphics[width=.49\textwidth]{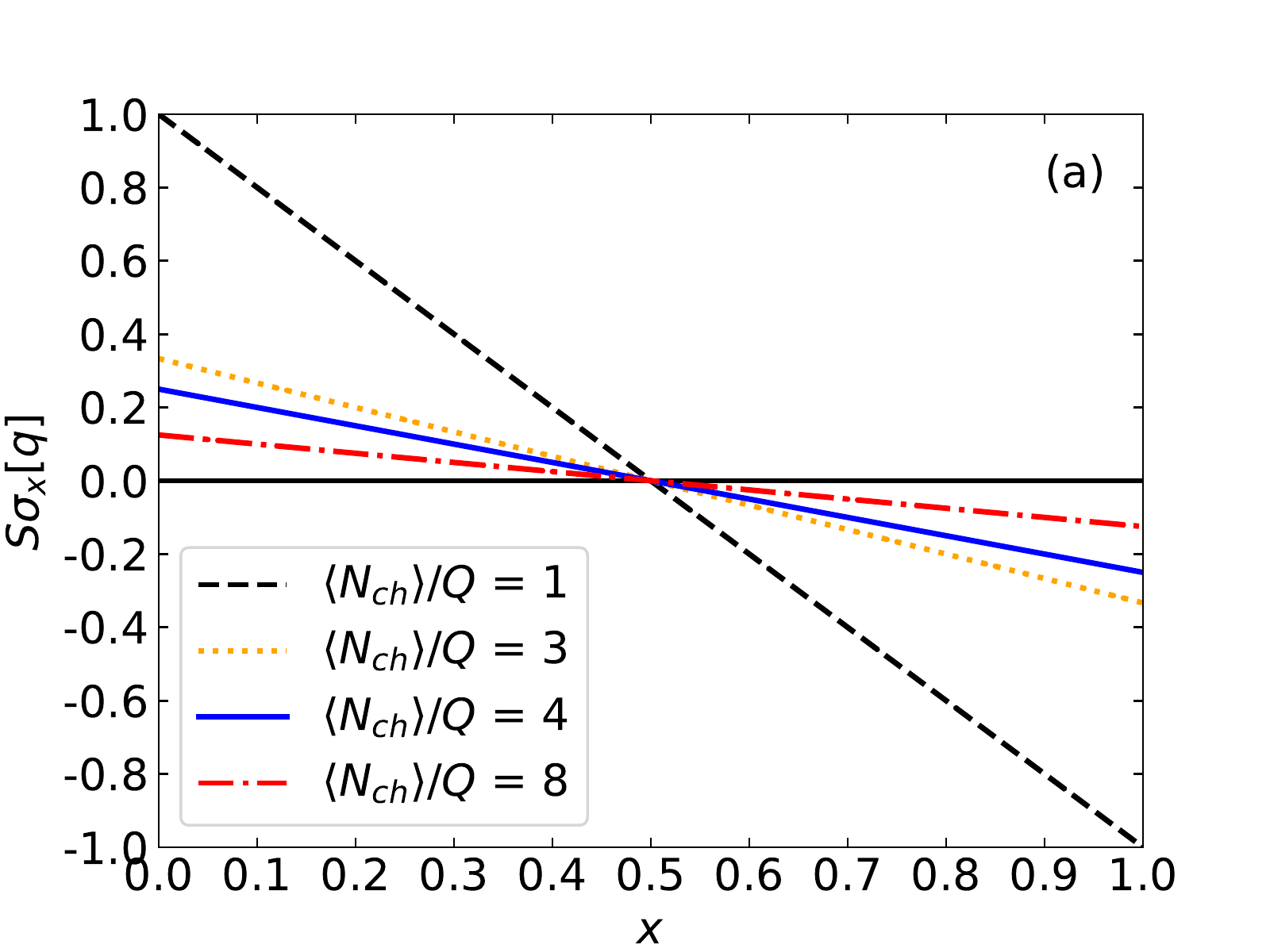}
\includegraphics[width=.49\textwidth]{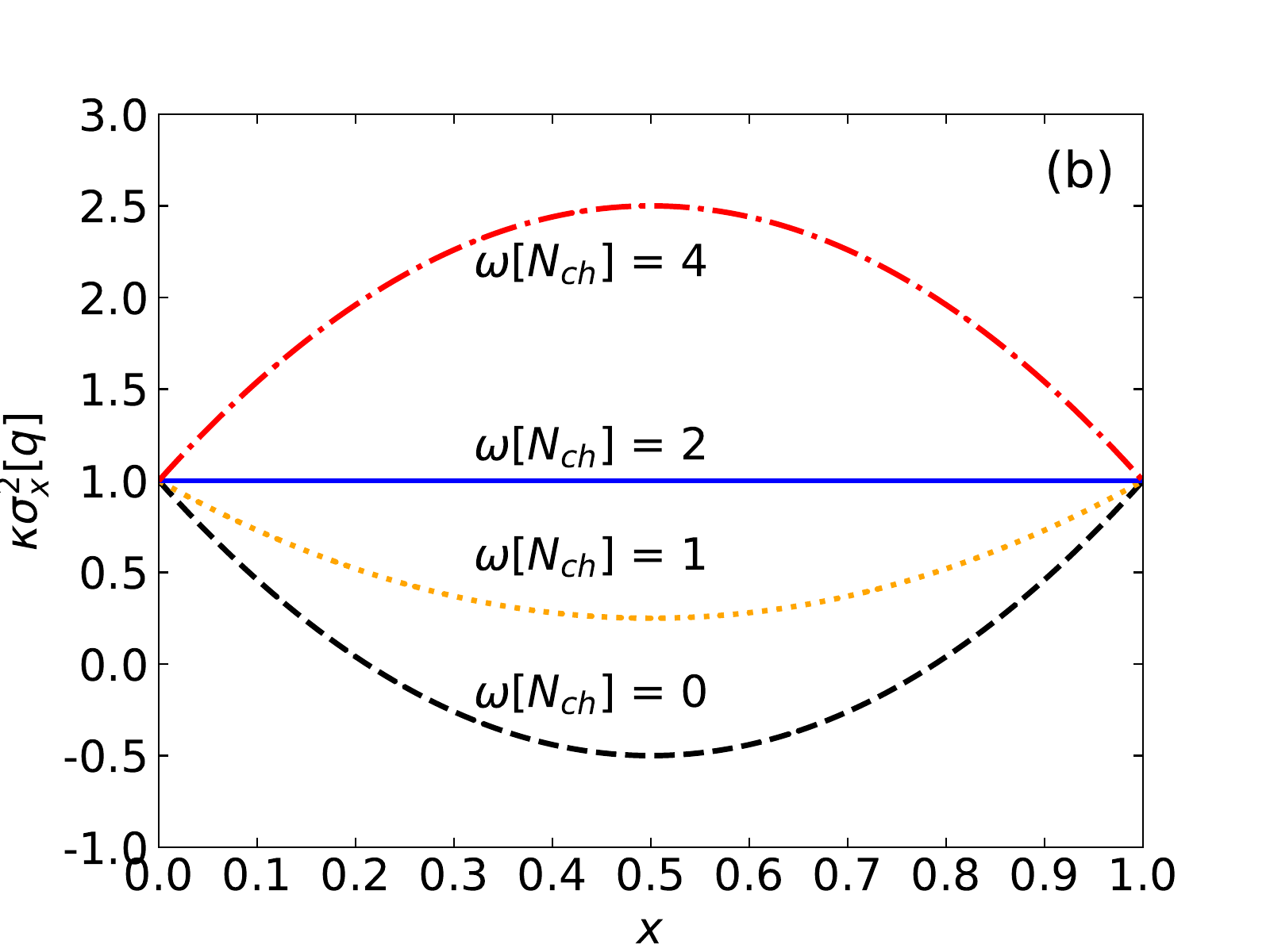}
\caption{The skewness (\ref{S[Q]}) (a) and kurtosis (\ref{kq}) (b) as  functions of the acceptance parameter $x$ at different values of $\langle N_{\rm ch}\rangle/Q  $ and $\omega[N_{\rm ch}]$.
}
\label{fig-kurt}
\end{figure*}

The net-charge skewness $S\sigma_x[q]$ (\ref{S[Q]}) and  kurtosis $\kappa \sigma_x^2[q]$ (\ref{kq}) depend, respectively, linearly and quadratically on the acceptance parameter $x$. These dependencies are shown in Fig.~\ref{fig-kurt} 
for different values of $\mean{N_{ch}}/Q$ and $\omega[N_{ch}]$. 
The ratio $Q / N_{ch}$ determines the slope of the $x$-dependence of the skewness. $S\sigma_x[q]$ equals zero at $x = 0.5$ for arbitrary values of $Q$ and $N_{ch}$.
The parabolic $x$-dependence of kurtosis $\kappa\sigma_x^2$ is determined by the value of  $\omega[N_{\rm ch}]$ only.
$\kappa\sigma_x^2$ (\ref{kq})  equals unity at $x = 0$ and $x = 1$ whereas the vertex of the parabolic dependence is located at $x=1/2$. These properties are independent of the  $\omega[N_{\rm ch}]$ value.
The value of $\omega[N_{\rm ch}]$ does define the concavity of the parabola:
it is convex for $\omega[N_{\rm ch}]< 2$, concave for $\omega[N_{\rm ch}]> 2$, and reduces to a horizontal line $\kappa\sigma_x^2 = 1$ for $\omega[N_{\rm ch}]= 2$~[see Fig.~\ref{fig-kurt} (b)].

We note that the data on $p+p$ reactions suggest that $\omega[N_{\rm ch}]$ is an increasing function of the collision energy with its values smaller than 2 at small collision energies and larger than 2 at large collision energies \cite{Konchakovski:2007ss}. Assuming $Q=2$, the values of $\langle N_{\rm ch}\rangle$ and $\omega[N_{\rm ch}]$ presented in Fig.~\ref{fig-kurt} correspond approximately to the $p+p$ data at $\sqrt{s}\cong 2$~GeV, 10~GeV, 20~GeV, and 100~GeV 
(see, e.g., Ref.~\cite{Konchakovski:2007ah}).
Similar arguments can be applied to baryon number fluctuations.
Note, however, that the fluctuations of $N_B+N_{\overline{B}}$ are essentially smaller than those of $N_++N_-$, i.e.,
$\omega[N_B+N_{\overline{B}}]\ll \omega[N_{\rm ch}]$. At the SPS and RHIC energies considered in this paper, one expects $\omega[N_B+N_{\overline{B}}] \le 1$.

For two statistically
correlated types of particles the cumulant generating function reads:
\eq{
&\ln[ F(k_+,k_-)]=\sum_{n,m= 1}^{\infty}\frac{\kappa_{n,m}[N_+,N_-]}{n!m!}(ik_+)^n (ik_-)^m \Rightarrow \\\label{most_general} &\ln[f_q(k|x_+,x_-)] = \sum_{n,m= 1}^{\infty}\frac{\kappa_{n,m}[N_+,N_-]}{n!m!}\phi[k|x_+]^n \phi[-k|x_-]^m,
}
as follows from Eq.~(\ref{fn}). Here $\kappa_{n,m}[N_+,N_-]$ are joint cumulants of $P\left(N_+,N_-\right)$. They obtain non-zero values if any correlation between positively and negatively charged particle is present. 

Then by taking respective derivatives the cumulants of the charge distribution can be obtained, see Eq.~(\ref{k_n_BAC})), which makes them linear functions of $\kappa_{n,m}[N_+,N_-]$. Note that Eq.~(\ref{most_general}) does not include factorial moments \cite{Bzdak:2012ab}.

We note that the binomial filter is also often used to correct for the detection efficiency~\cite{Bzdak:2012ab,Luo:2014rea,Kitazawa:2016awu}, i.e. for the fact that the measurable particles within the acceptance are not detected with a 100\% probability.
The detection efficiency typically is dependent on the number of measured particles and their momenta. 
Therefore, the simple binomial filter may not be sufficient and more involved local and multiplicity-dependent efficiency corrections have been considered instead~\cite{Bzdak:2013pha,Bzdak:2016qdc}.
As published experimental data are typically corrected for the detection efficiency, we do not consider efficiency corrections in the present work.

\section{(Grand-)Canonical statistical mechanics}
\label{sec-poisson}

In this section, we analyze a couple of common full space multiplicity distributions in statistical mechanics.
We consider grand-canonical and canonical distributions of relativistic particles, which represents two useful baselines in the context of heavy-ion collisions.

The grand-canonical multiplicity distribution of non-interacting Maxwell-Boltzmann particles in equilibrium is given by a Poisson distribution.
In a relativistic case studied here,
the joint probability distribution of the numbers of particles $N_+$ and antiparticles $N_-$ is given by a product of two Poisson distributions:
\begin{equation}
{\cal P_{GCE}}(N_{+},N_{-}) ~ \sim ~  \frac{z_+
^{N_{+}}}{N_{+}!}\frac{z_-^{N_{-}}}{N_{-}!}~. \label{poisson}
\end{equation}
Here the quantities $z_+$ and $z_-$ are defined as
\eq{\label{z}
z_\pm = \exp\left(\pm\,\frac{ \mu_Q}{T}\right)\, \frac{gV}{2\pi^2}\int_0^\infty k^2dk\exp\left[-\frac{\sqrt{k^2+m^2}}{T}\right] \equiv \exp\left(\pm\,\frac{ \mu_Q}{T}\right)~z~,
}
where $V$, $\mu_Q$, and $T$ are, respectively, the system volume, charge chemical potential, and temperature. 
$g$ and $m$ are the particle degeneracy factor and mass. 
The chemical potential regulates the mean net number of particles and antiparticles, $\langle Q \rangle = \mean{N_+} - \mean{N_-}$.
A generalization to a system with multiple particle species carrying a conserved charge $Q$ is achieved by simply adding contributions of these extra species to Eq.~\eqref{z}.

One finds $\omega[N_+]=\omega[N_-]=\omega[N_{\rm ch}]=1$. 
These fluctuation measures are shown by the horizontal dashed line in Fig.~\ref{fig-bessel}~(a). 
The charge distribution $P_{\rm Sk}(Q)$   corresponds then to the so-called Skellam distribution \cite{10.2307/2980526}. The cumulants $k_n[Q]$ for the Skellam distribution $P_{\rm Sk}(Q)$ can be easily found as
\eq{\label{skelam-cum}
k_n[Q]=z_+ + (- 1)^nz_-~.
}
This gives
\eq{\label{skellam-skew}
S\sigma[Q]= \frac{z_+-z_-}{z_++z_-}= \left[1+\left(\frac{2z}{\langle Q\rangle }\right)^2\right]^{-1/2}~,~~~~~~ \kappa\sigma^2[Q]=1~.
}
The skewness $S\sigma[Q]$ of the Skellam distribution, 
given by Eq.~(\ref{skellam-skew}), is shown in Fig.~\ref{fig-bessel} (b) by the dashed line for $\langle Q\rangle =2$.
For $x_-=x_+\equiv x$ 
one finds $\kappa_n[q]=x \kappa_n[Q]$ for cumulants $\kappa_n[q]$ evaluated in an $x$-acceptance. 
This implies $k_m[q]/k_n[q]=k_m[Q]/k_n[Q]$, meaning that all intensive fluctuation measures calculated via the BAC for  an arbitrary fixed $x<1$ are equal to their values in the full phase space ($x=1$).

\begin{figure*}[t]
\includegraphics[width=.49\textwidth]{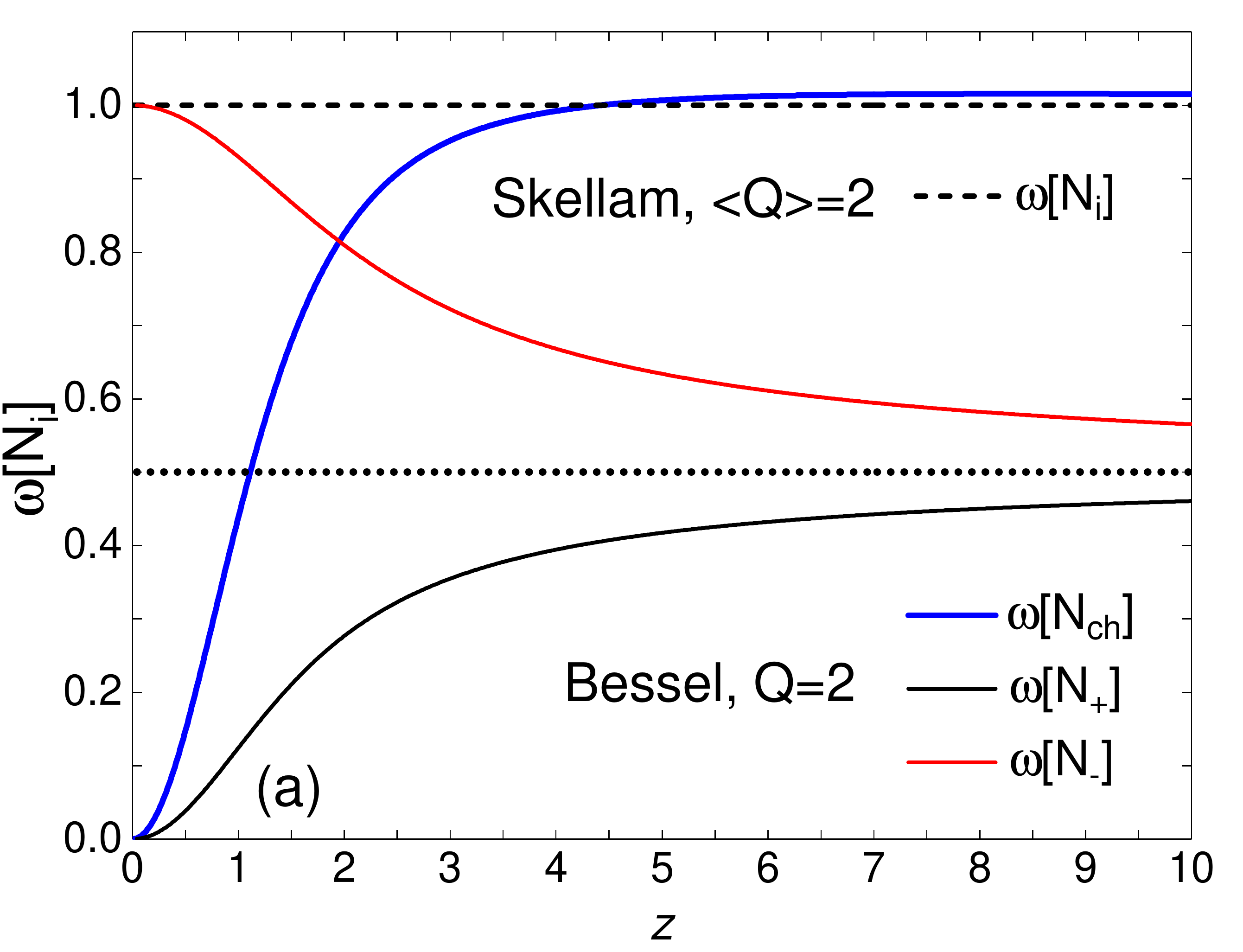}
    \includegraphics[width=.49\textwidth]{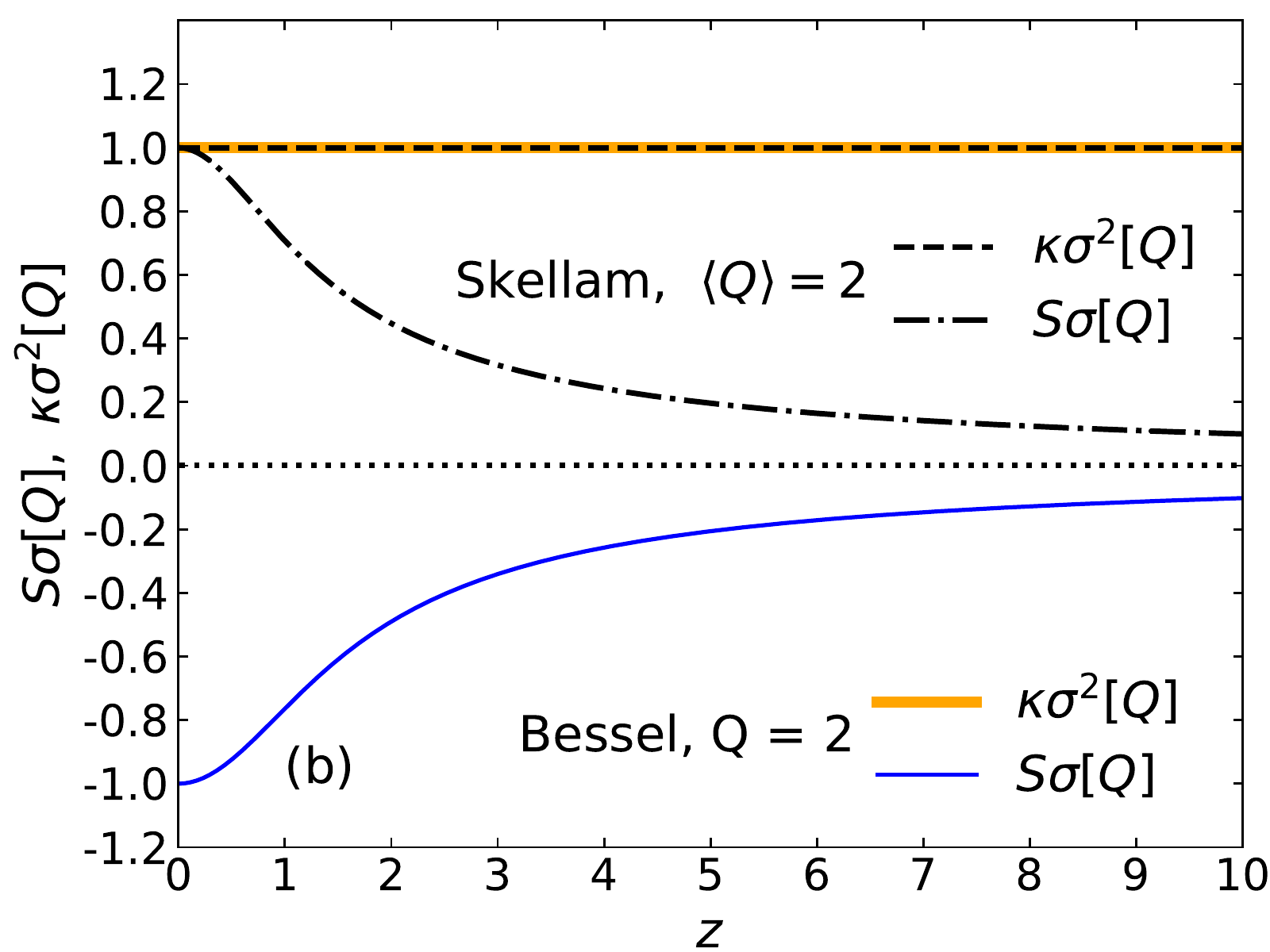}
    \includegraphics[width=.49\textwidth]{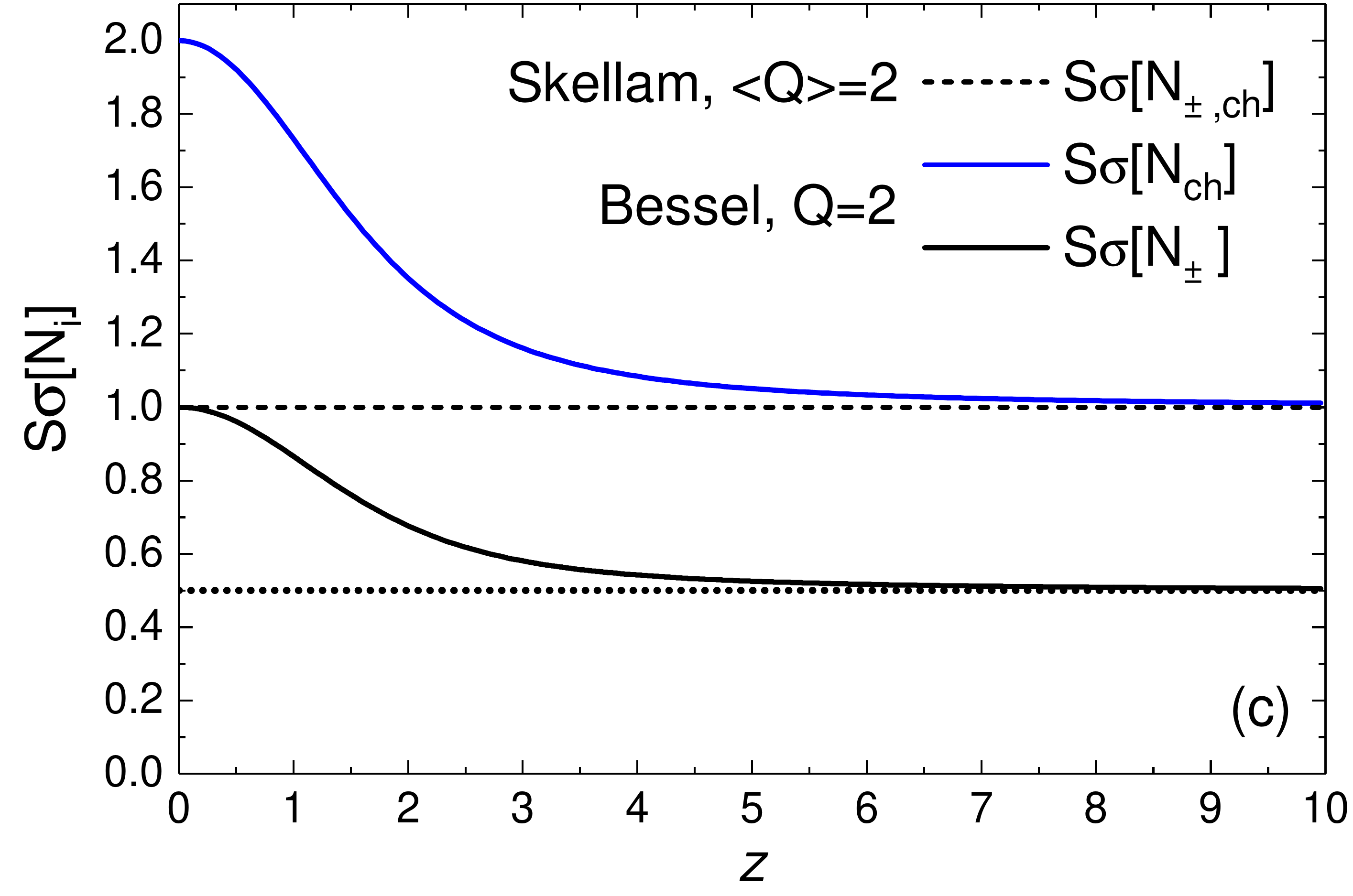}
    \includegraphics[width=.49\textwidth]{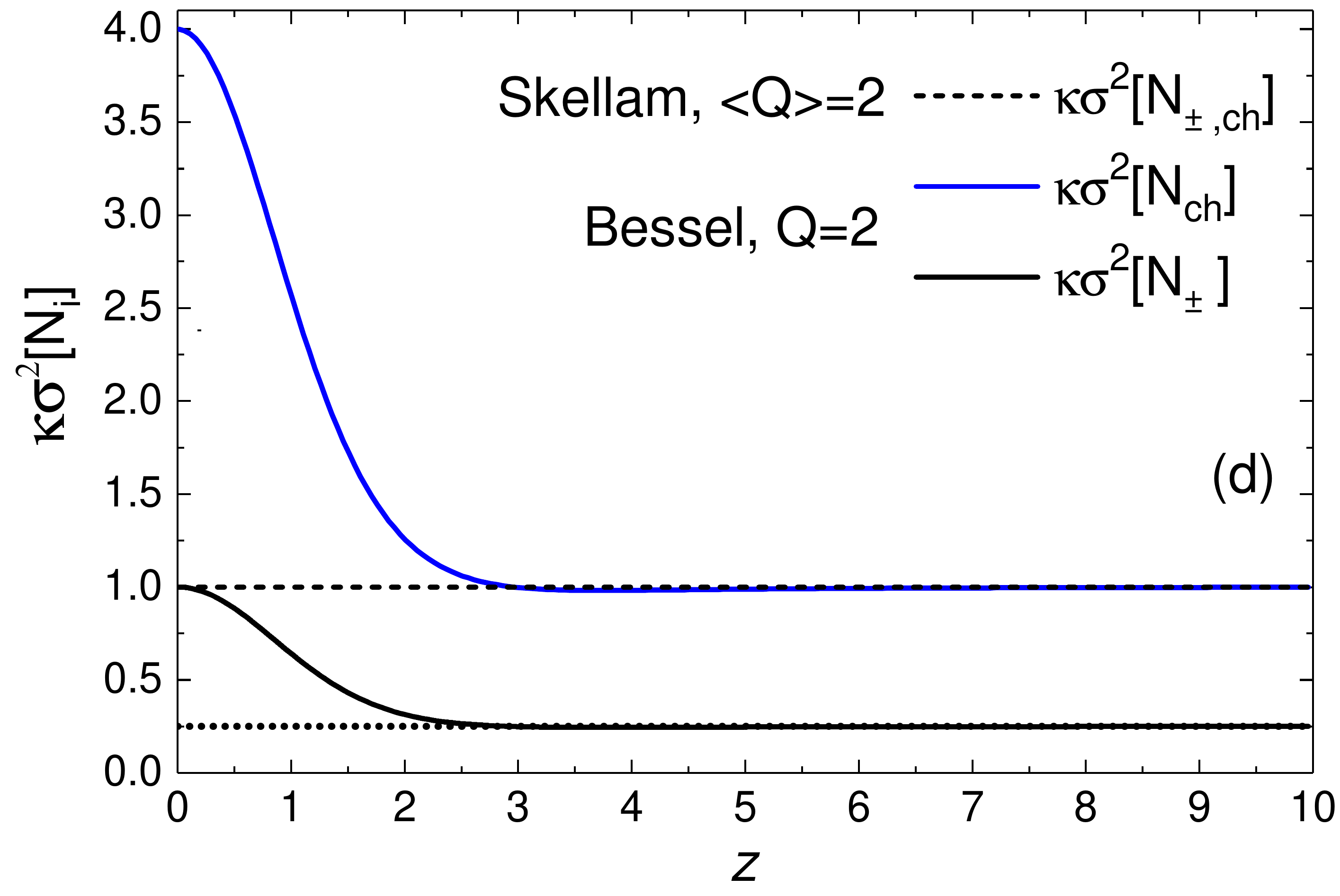}
\caption{ The scaled variances $\omega[N_i]$ (with $i=+,-,{\rm ch}$) (a), skewness $S\sigma[Q]$ and kurtosis $\kappa\sigma^2[Q]$ (b), $S\sigma[N_i]$ (c),  $\kappa\sigma^2[N_i]$ (d)  for the Skellam distribution (\ref{poisson}) with $\langle Q\rangle =2$ (dashed line) and for the Bessel distribution (\ref{bessel}) with $Q=2$ (solid lines) are shown as functions of $z$.
}  \label{fig-bessel}
\end{figure*}

The canonical ensemble entails global charge conservation in all microscopic states of a statistical system. 
The scaled variances for fluctuations of $N_+$, $N_-$,
and $N_{\rm ch}$ in an ideal classical gas within the canonical ensemble were considered in Refs.~\cite{Begun:2004gs,Begun:2004zb}, while similar questions related to baryon number conservation were discussed in Ref.~\cite{Braun-Munzinger:2018yru}. 
The particle number  distribution ${\cal P}(N_+,N_-)$  corresponds to a two-Poisson distribution with a fixed difference:
\begin{equation}
{\cal P_{CE}}(N_{+},N_{-}) ~\sim ~  \frac{z^{N_{+}}}{N_{+}!}\frac{z^{N_{-}}}{N_{-}!}~\delta(N_{+}-N_{-}-Q)~, \label{bessel}
\end{equation}
where the parameter $z$ is defined in Eq.~(\ref{z}) and 
is proportional to the volume of statistical system.
It follows from Eq.~\eqref{bessel} that numbers $N_{\pm}$ and $N_{ch} = N_+ + N_-$ are both described by various forms of the Bessel distribution \cite{Yuan2000}. Their characteristic functions are the following:
\eq{\label{FN}
F_{N_{\pm}}(k) & \equiv \sum_{N_+,N_-=0}^{\infty}e^{ikN_{\pm}}{\cal P}(N_+,N_-)= e^{\pm iQ\frac{k}{2}}\frac{I_{Q}(2ze^{i\frac{k}{2}})}{I_{Q}(2z)}~,\\
F_{N_{ch}}(k) & \equiv \sum_{N_+,N_-=0}^{\infty}e^{ikN_{\rm ch}}{\cal P}(N_+,N_-) = \frac{I_{Q}(2ze^{ik})}{I_{Q}(2z)}~.\label{Fch}
}
Here $I_{Q}$ is the modified Bessel function of the first kind. The expressions (\ref{FN}) and (\ref{Fch}) depend on two parameters: $Q$ and $z$. 
The cumulants $\kappa_l[N_\pm]$ and $\kappa_l[N_{\rm ch}]$ are obtained by taking derivatives of the corresponding cumulant generating functions $\ln(F_{N_\pm})$
and $\ln(F_{N_{\rm ch}})$.
The mean values and the scaled variances read
\eq{
\langle N_{ch} \rangle & = z\left(a_{+}+a_{-}\right), \\
\langle N_{\pm} \rangle & = \frac{\langle N_{ch} \rangle \pm Q}{2}, \\
\omega[N_{ch}] & = 1 + z\left[\frac{2+b_{+}+b_{-}}{a_+ + a_-}-(a_{+}+a_{-})\right], \\
\omega[N_{\pm}] & = 1 - z\left[a_{\pm}-\frac{b_{\pm}}{a_{\pm}}\right],
}
where $a_{\pm}= I_{Q\pm 1}[2z]/I_{Q}[2z]$, $b_{\pm}= I_{Q\pm 2}[2z]/I_{Q}[2z]$.

Simplied expressions can be obtained in certain limits.
For large systems, $z \gg Q$ and $z \gg 1$, one has  $\kappa_{l}[N_{\pm}] \approx (2z)/2^{l}$ and $\kappa_{l}[N_{ch}]= 2^{l}k_{l}[N_{\pm}]$. 
Explicit expressions for means and variances in this limit read
\eq{
\langle N_+\rangle \cong \langle N_- \rangle \cong ~\frac{1}{2}\langle N_{\rm ch}\rangle\cong z~,~~~~
 \omega[N_\pm]\cong  \frac{1}{2}+\frac{1}{8z} \mp \frac{Q}{4z}~, ~~~~~~  
 \omega[N_{\rm ch}]\cong 1 + \frac{1}{4z}~.
}

For $z \ll \sqrt{Q+1}$ the cumulant generating functions of $N_{\pm}$ and $N_{ch}$ read
\begin{equation}
\ln[F_{N_{\pm}}(k)] = ikQ\delta_{\pm,+} + \frac{z^{2}}{[Q+1]}\sum_{l=1}^{\infty}\frac{(ik)^l}{l!},~~~~~~
\ln[F_{N_{ch}}(k)] = ikQ + \frac{z^{2}}{[Q+1]}\sum_{l=1}^{\infty}\frac{(2ik)^l}{l!},
\end{equation}
and $\kappa_{l}[N_{\pm}] = z^{2}/[Q+1]$, $\kappa_{l}[N_{ch}] = 2^{l}k_{l}[N_{\pm}]$, $l>1$. 
Using the above equations one calculates the asymptotic behavior 
at $z\ll \sqrt{Q+1}$ :
\eq{
& \langle N_+\rangle \cong Q, ~~~~~~~ \langle N_- \rangle \cong \frac{z^2}{Q+1}, ~~~~~~\langle N_{\rm ch}\rangle  = Q~,\\  
 & \omega[N_+] \cong 1 - \frac{z^2}{(Q+1)(Q+2)},~ ~~~~~\omega[N_-]\cong \frac{z^2}{Q(Q+1)}~,~~~~~
   ~\omega[N_{\rm ch}]= \frac{4z^2}{Q(Q+1)}~.
}
Figure~\ref{fig-bessel} (a) presents the $z$-dependence of scaled variances $\omega[N_{\rm ch}]$ and $\omega[N_{\pm}]$ for the Bessel distribution (\ref{bessel}) with $Q = 2$.

The skewness and kurtosis for $N_{\pm}$ and $N_{ch}$ fluctuations are presented in Fig.~\ref{fig-bessel} (c) and (d), respectively. Relation $\kappa_{l}[N_{ch}] = 2^{l}k_{l}[N_{\pm}]$ has been verified to hold for all values of $z$, as can be seen from Fig.~(\ref{fig-bessel}) (a),(c), and (d). 
Higher-order cumulant ratios of charged particle number fluctuations are enhanced by the exact charge conservation. 
Moreover, the $N_{ch}$ cumulants are not equal to the sum of cumulants of negatively and positively charged particles. 
We attribute this to a presence of multiparticle correlations induced by the exact charge conservation. 

The net-charge $Q$ is conserved globally and does not fluctuate in the full space, i.e., in the limit $x \to 1$.
The scaled variance of net-charge fluctuations in full space is, therefore, vanishing: $\omega[Q] = 0$.
The skewness and kurtosis of $Q$, on the other hand, attain finite values in the limit $x \to 1$, which follow from Eqs.~(\ref{w_x[Q]})-(\ref{kq}): 
\eq{
S\sigma[Q] = -\frac{Q}{\langle N_{ch}\rangle},~~~~~~\kappa\sigma^2[Q] = 1~.
}
The behavior of $S\sigma[Q]$ and $\kappa\sigma^2[Q]$ for the Bessel distribution with $Q = 2$ is shown in Fig.~\ref{fig-bessel} (b). 

The above results illustrate the non-trivial behavior of fluctuation measures of positively or negatively charged particle numbers that arise due to the exact conservation of the net charge. This behavior is present already for fluctuations in the full phase space, $x = 1$.
The BAC expressions, Eqs.~(\ref{w-x})-(\ref{k_x}), for single charge and Eqs.~(\ref{k1x1x2})-(\ref{k4x1x2}) for net charge, allow then to obtain the corresponding behavior in a finite acceptance, $x < 1$.
It should be pointed out, however, that the results of this Section are obtained for an idealized system. For example, effects of resonance decays have been neglected here. These decays produce an additional source of correlations.
More generally, one should also consider a simultaneous conservation of all three conserved charges, baryon number, electric charge, and strangeness, rather than of just a single conserved charge.
These effects are incorporated in UrQMD transport model studies that we present below.

\section{UrQMD simulations of $p+p$ reactions}
\label{sec-urqmd}

Transport simulations can provide useful information about the acceptance dependence of fluctuations, and test accuracy of the BAC in various setups.
Earlier, to study the net charge fluctuations within transport model the hadronic matter simulations in a box with periodic boundary conditions were used ~\cite{Petersen:2015pcy}. 
Cuts in coordinate space were applied, i.e., it was assumed that the detection of particles takes place only inside the subsystem with volume $v=x V$, where $V$ denotes the total volume of a box with periodic boundary conditions and $x$ is the acceptance parameter. 
It was shown that the acceptance dependencies of the skewness and kurtosis of net charge fluctuations in such a system do satisfy the BAC predictions.
Because the multiplicity distribution ${\cal P}(N_+,N_-)$ for hadrons inside the box within transport models appears to be close to the Bessel distribution~(\ref{bessel}),
only convex downward curves for $\kurt[Q]$ where obtained. 

Actual high-energy collision experiments measure the momenta of final state particles rather than coordinates.
Therefore, the BAC should be considered in the momentum space.
In this section the BAC predictions for fluctuations of particle numbers, as well as of conserved charges $B$ and $Q$, are compared with results of the UrQMD transport model~\cite{Bass:1998ca,Bleicher:1999xi} simulations of inelastic $p+p$ reactions.
UrQMD is an event generator producing a list of hadrons and their momenta in the final state of the collision. The generator satisfies the exact conservation of energy-momentum and of all the QCD conserved charges.
It also naturally incorporates correlations between particles emerging from resonance decays and string fragmentations. Acceptance cuts in the momentum space can be applied straightforwardly, making UrQMD suitable for direct comparisons to data. This is in contrast to statistical-thermal models where additional assumptions are needed, the BAC being one such possible assumption. The measured hadron multiplicities and momentum spectra calculated within UrQMD simulations are usually in a fair agreement with available experimental data.
All in all, this makes UrQMD a useful tool to analyze the behavior of fluctuations in various acceptance windows, and to test the performance of the BAC in various setups.

Here we analyze inelastic $p+p$ collisions at an SPS energy of $\sqrt{s} = 6.3$~GeV as well as at one of the RHIC energies, namely $\sqrt{s} = 62.4$~GeV.
In $p+p$ collisions the electric charge and the net baryon number are equal to $Q=B=2$ and do not fluctuate in the full phase space. 
We shall analyze in some detail the acceptance dependence of fluctuations of positively and negatively charged hadron multiplicities, as well as of (net) baryon number and net charge.

\subsection{Rapidity window dependence of acceptance parameters}

\begin{figure*}[h]
\includegraphics[width=.49\textwidth]{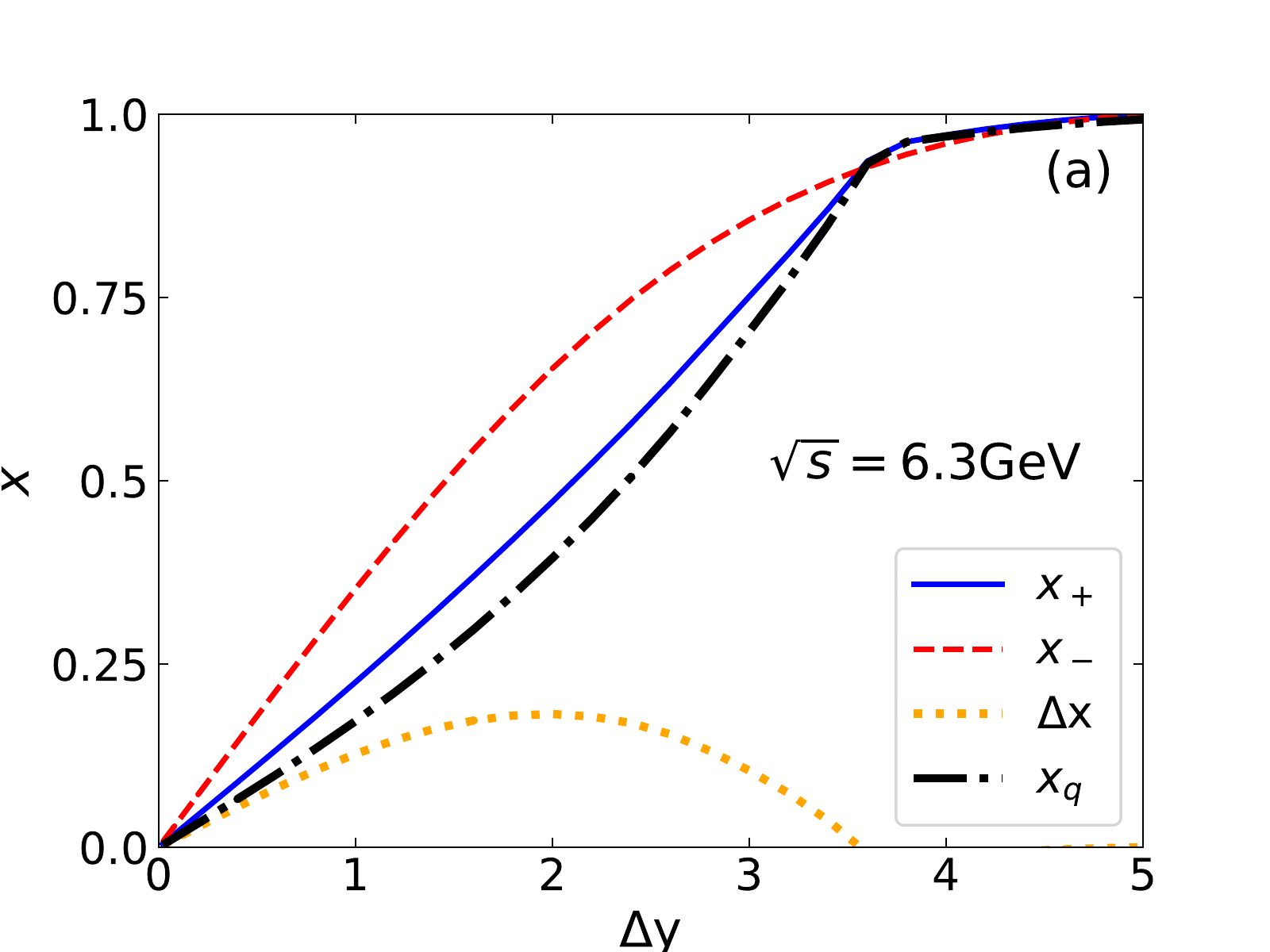}
\includegraphics[width=.49\textwidth]{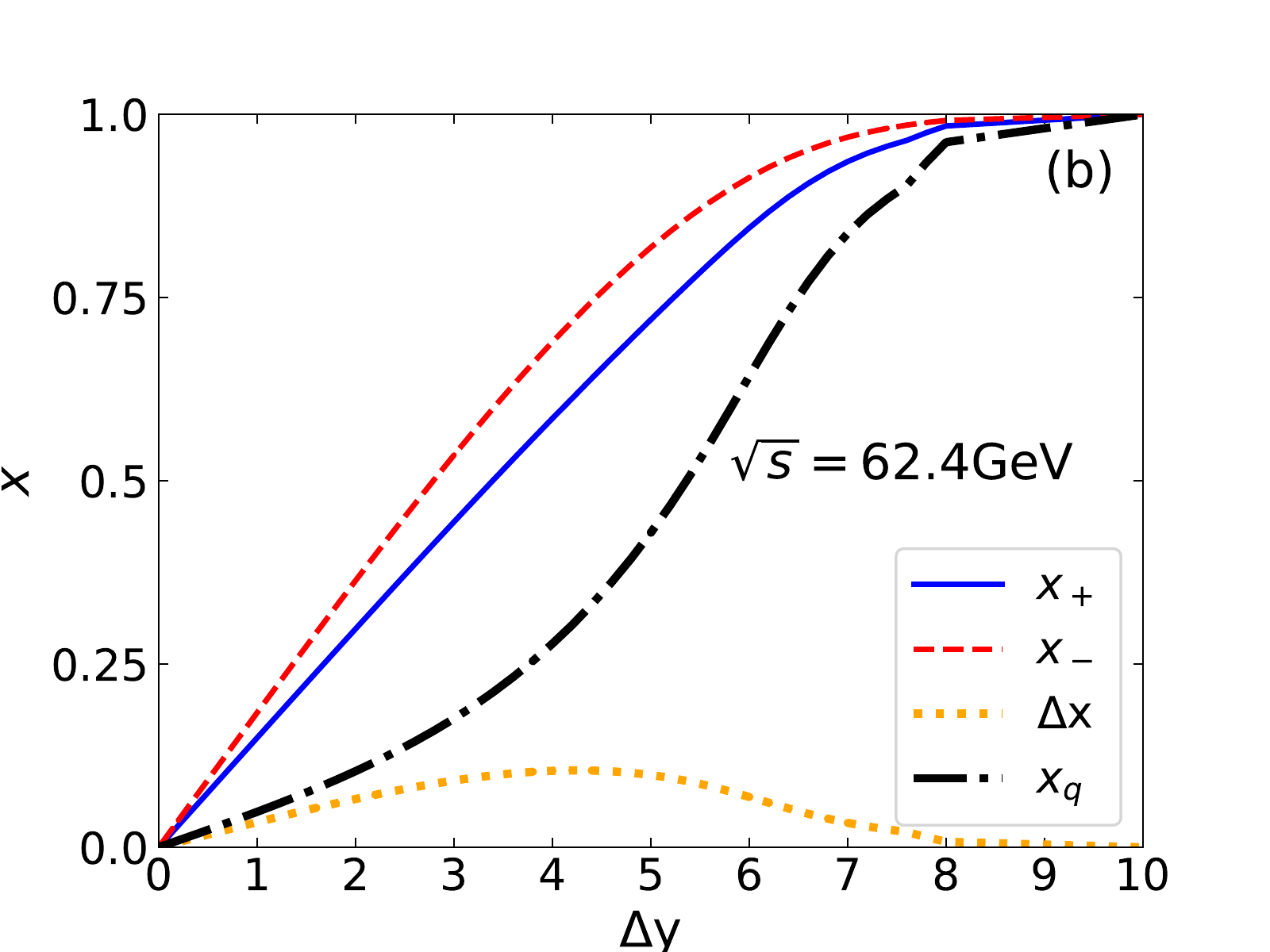}
\includegraphics[width=.49\textwidth]{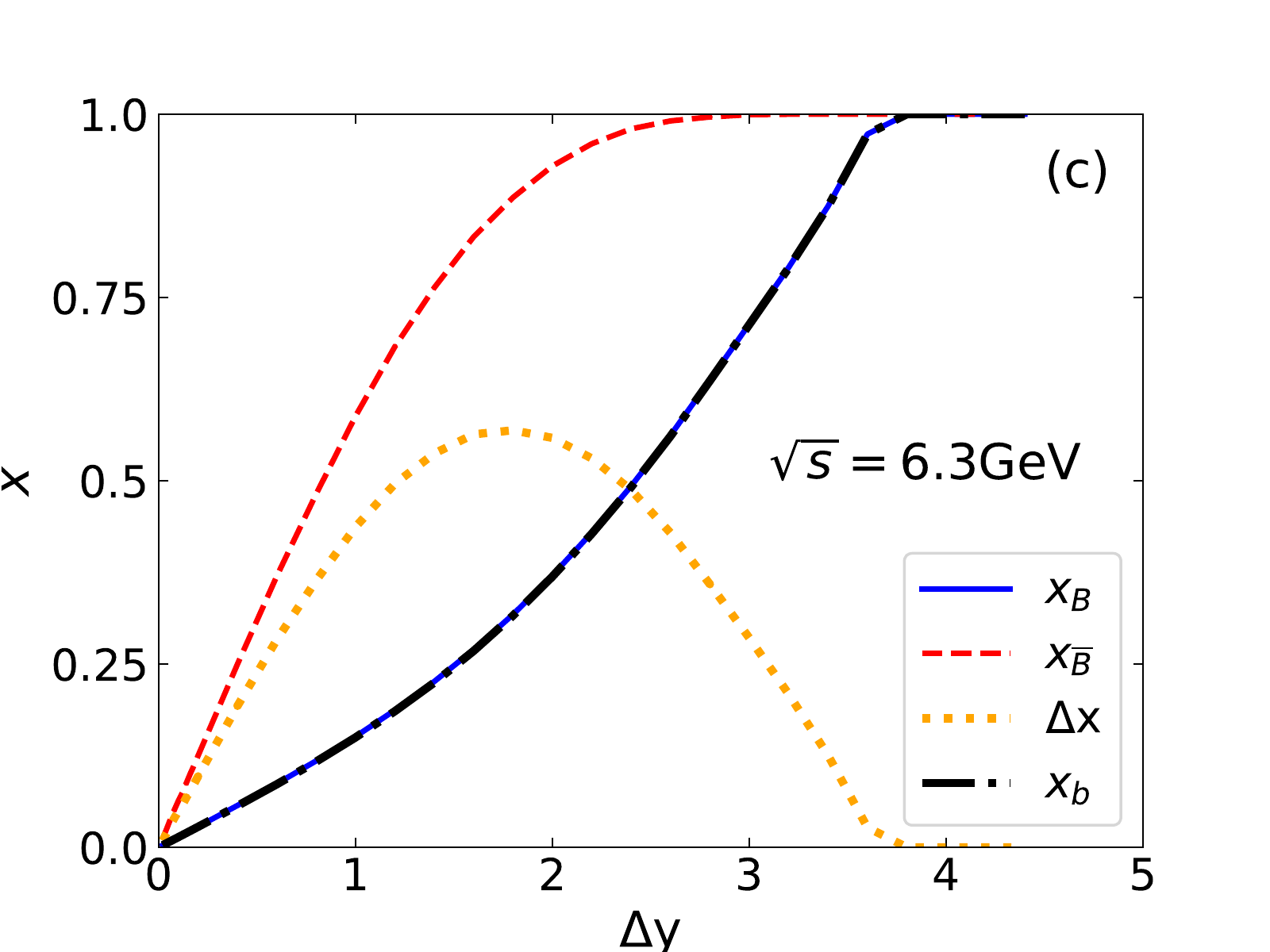}
\includegraphics[width=.49\textwidth]{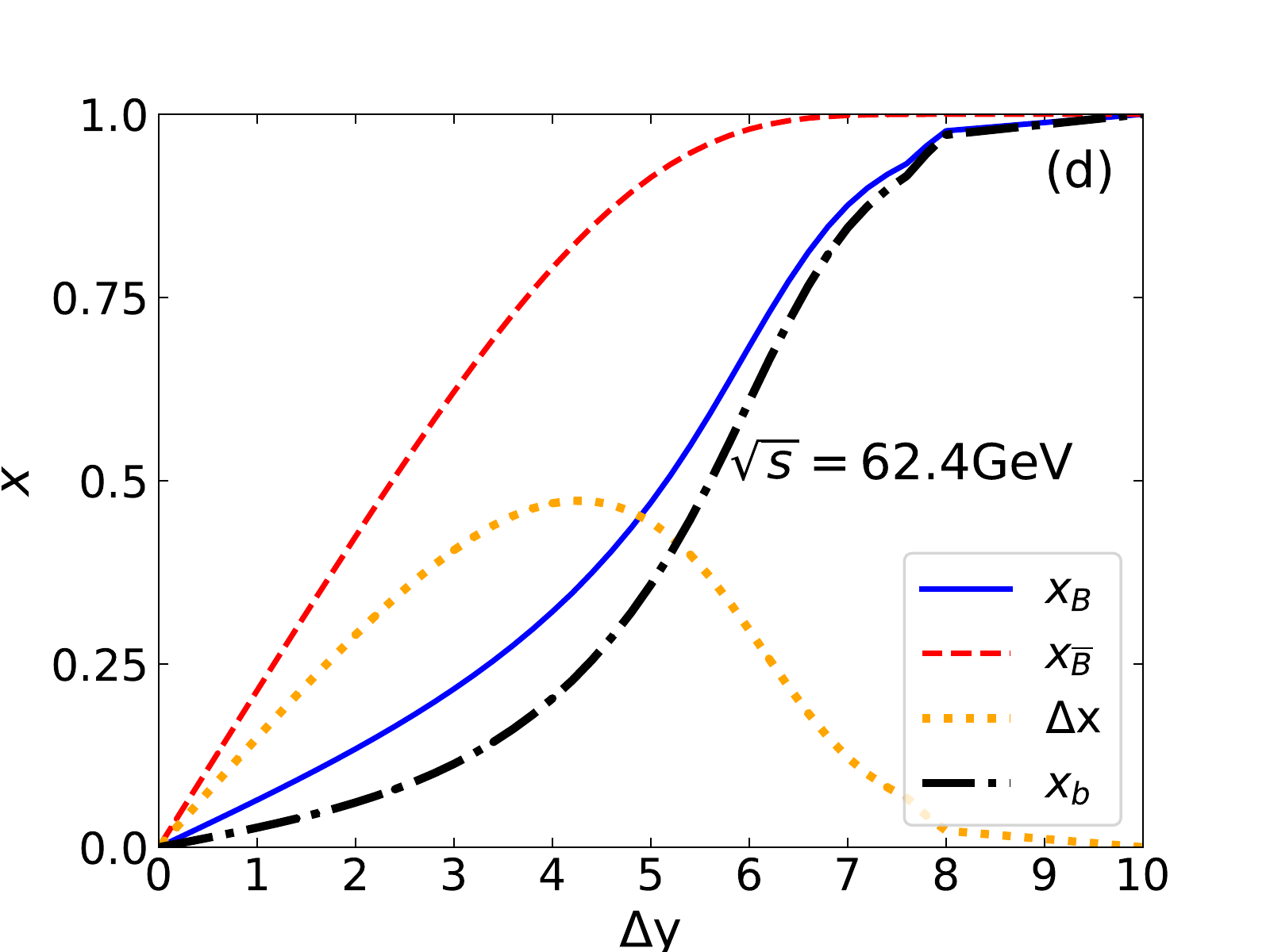}
\caption{\label{vol_versus_rap}
The acceptance $x_i$-parameters as functions of the rapidity interval $\Delta y$
calculated in the UrQMD model. 
The orange lines present the difference $\Delta x=x_--x_+$ or $\Delta x=x_{\bar{B}}-x_B$. The black dash-dotted lines present acceptance of  (a),(b) the net electric charge or (c),(d) of the net baryon number.
}
\end{figure*}

The same value of BAC $x$-parameter corresponds to quite different regions in the momentum space at different collision energies.
To be definite, we chose the acceptance region as a $p_T$-integrated finite rapidity interval $-\Delta y/2 \le y \le \Delta y/2$
in the center of mass of the system.
Any particle in this rapidity interval is assumed to be accepted with 100\% probability, therefore the BAC parameter
\eq{\label{xi}
x_i=  \frac{\int\limits_{-\Delta y/2}^{\Delta y/2}dy~\frac{dN_i}{dy}}{ \int\limits_{-\infty}^{\infty}dy~\frac{dN_i}{dy}}~\equiv~\frac{\langle n_i\rangle}{\langle N_i\rangle}~
} 
is simply the ratio between the mean number of particles in the acceptance relative to the one in the full phase space.

\begin{figure*}
 \includegraphics[width=.49\textwidth]{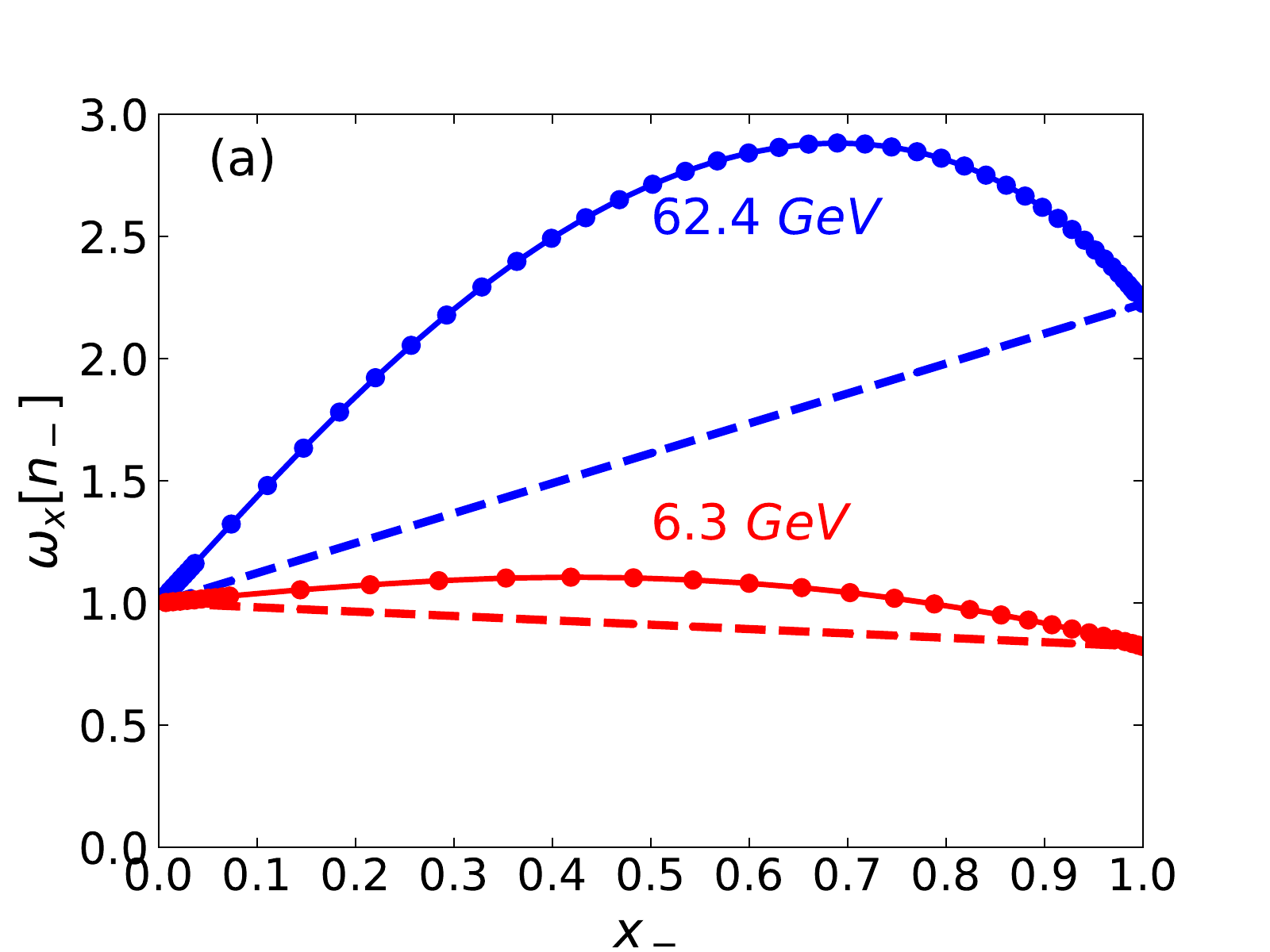}
\includegraphics[width=.49\textwidth]{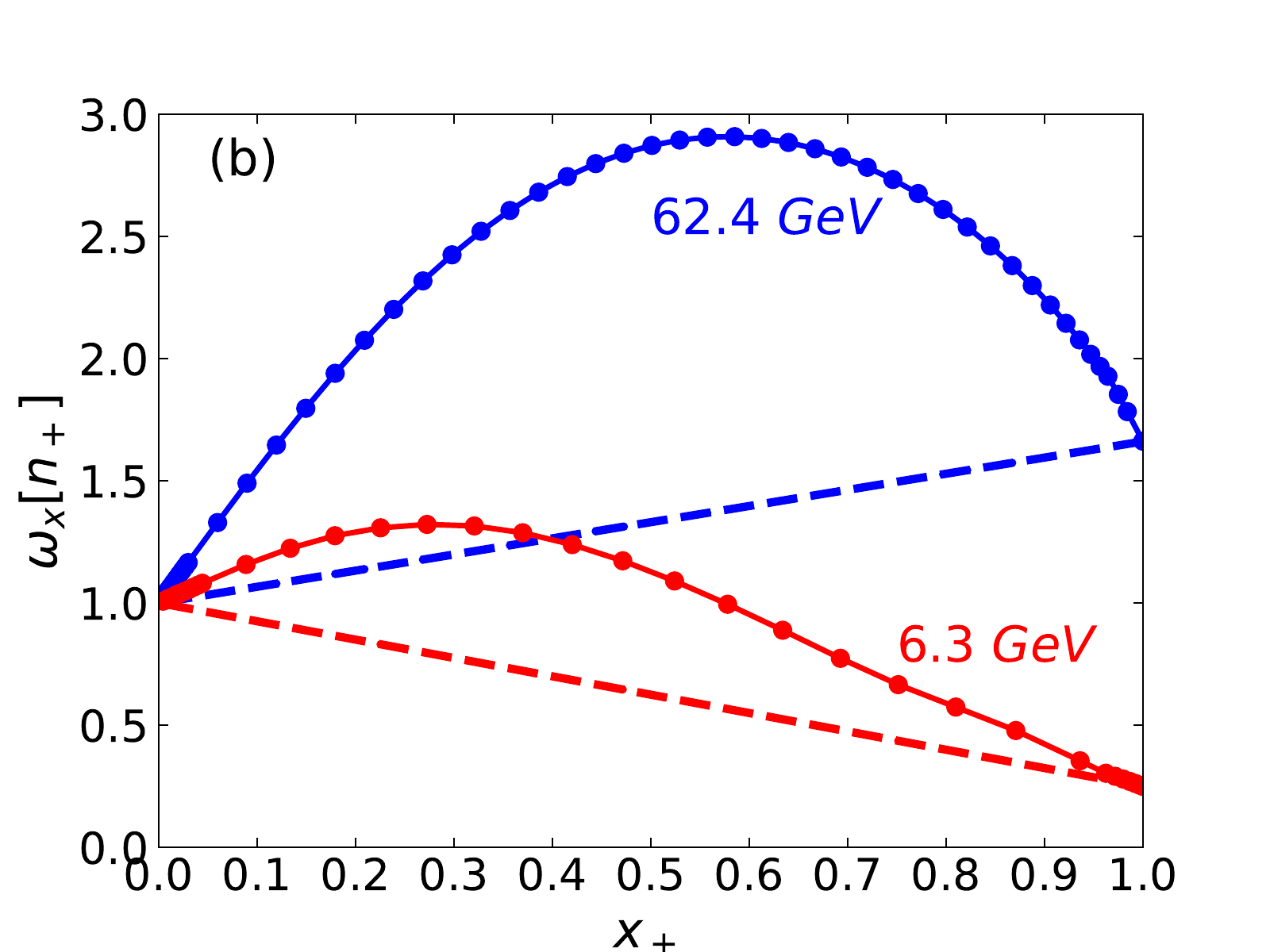}
\includegraphics[width=.49\textwidth]{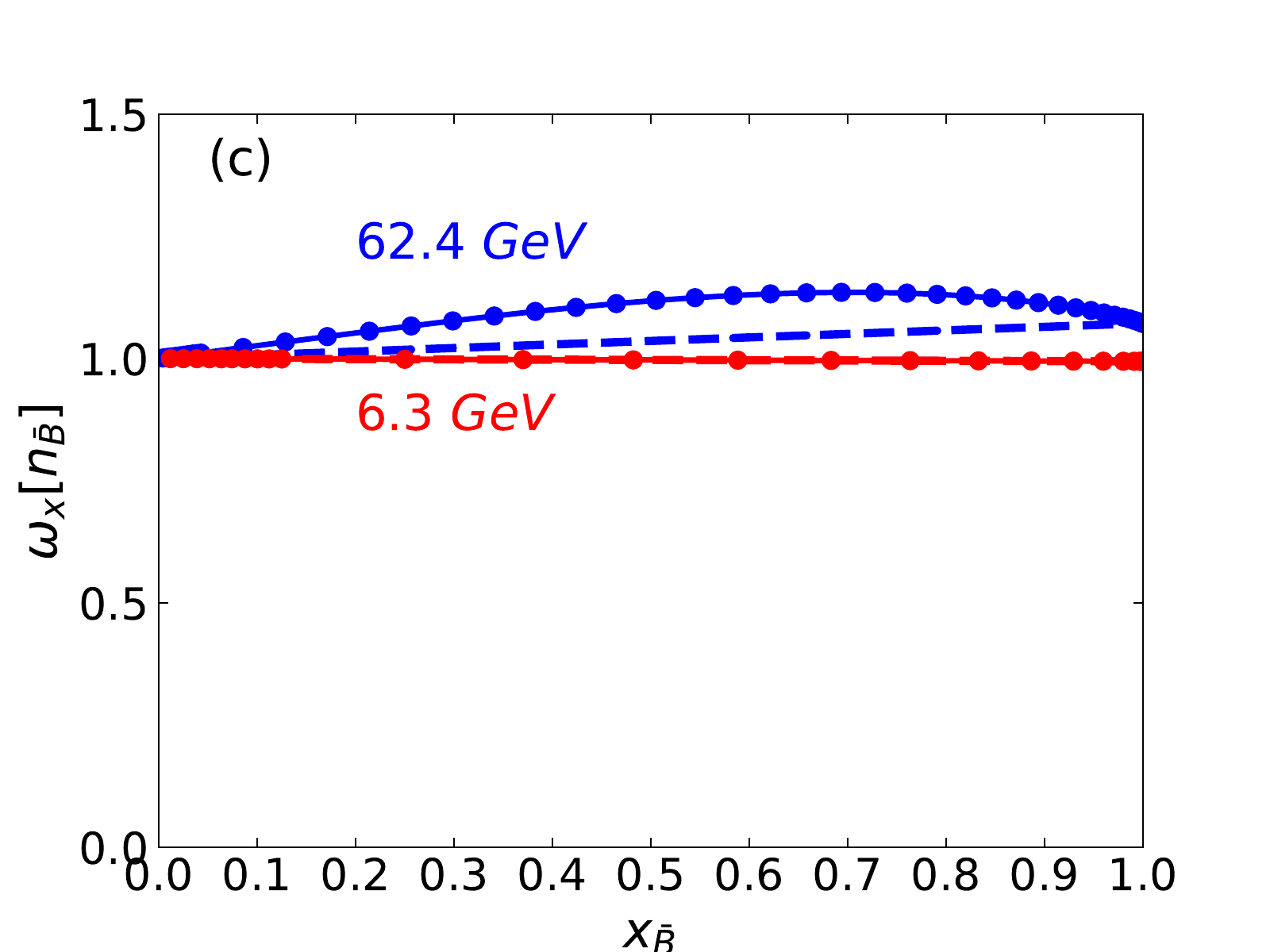}
\includegraphics[width=.49\textwidth]{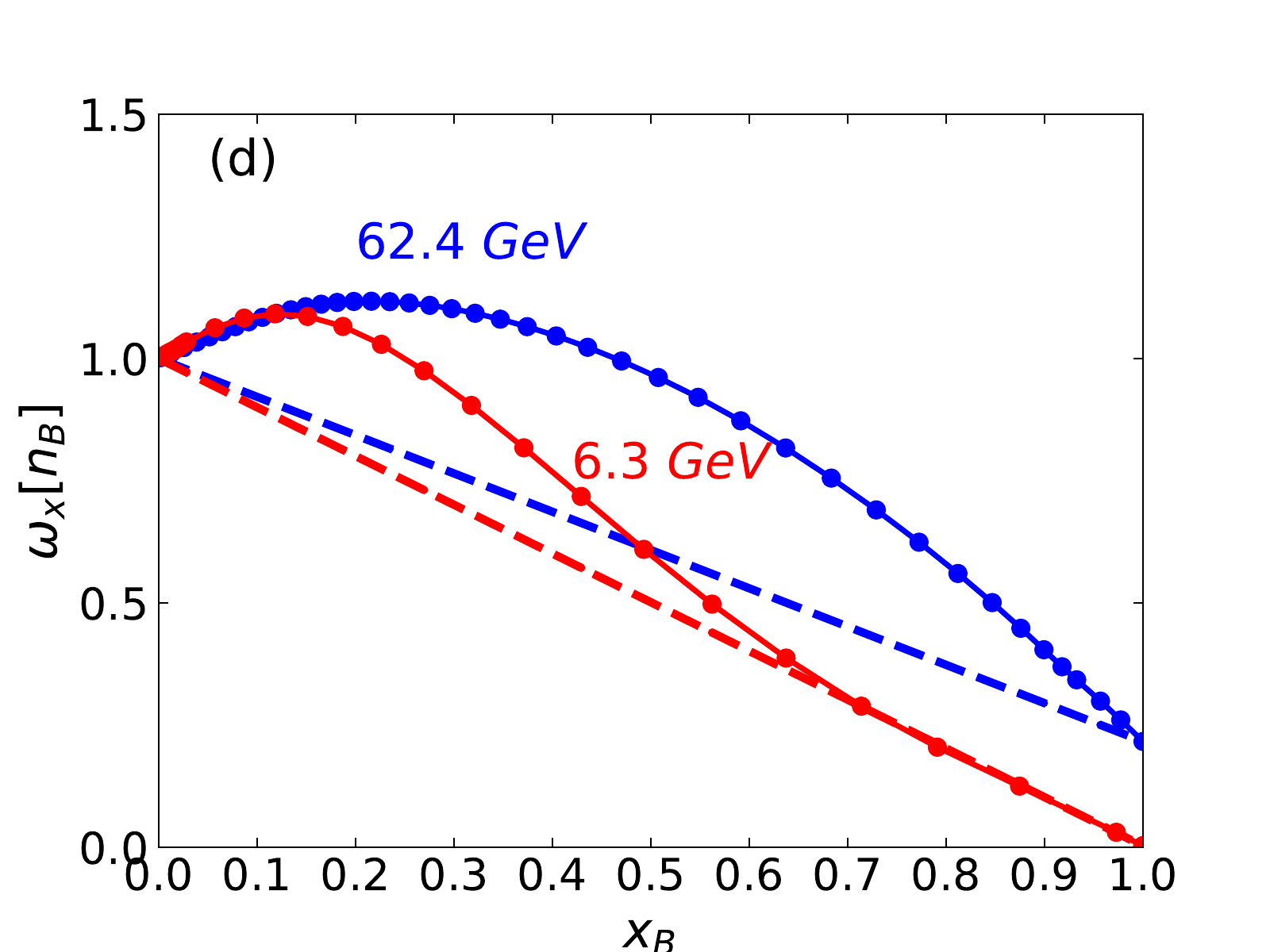}
\caption{\label{fig-n-}
Scaled variances $\omega_x[n_i]$ 
as functions of $x_i$.
Blue and red points are calculated for $p+p$ collisions in the UrQMD model and represent, respectively, energies of  $\sqrt{s}=6.3$ and $\sqrt{s}=62.4$~GeV. The BAC results  (\ref{w-x})-(\ref{k_x}) are shown by dashed lines.
}
\end{figure*}

\begin{figure*}
\includegraphics[width=.49\textwidth]{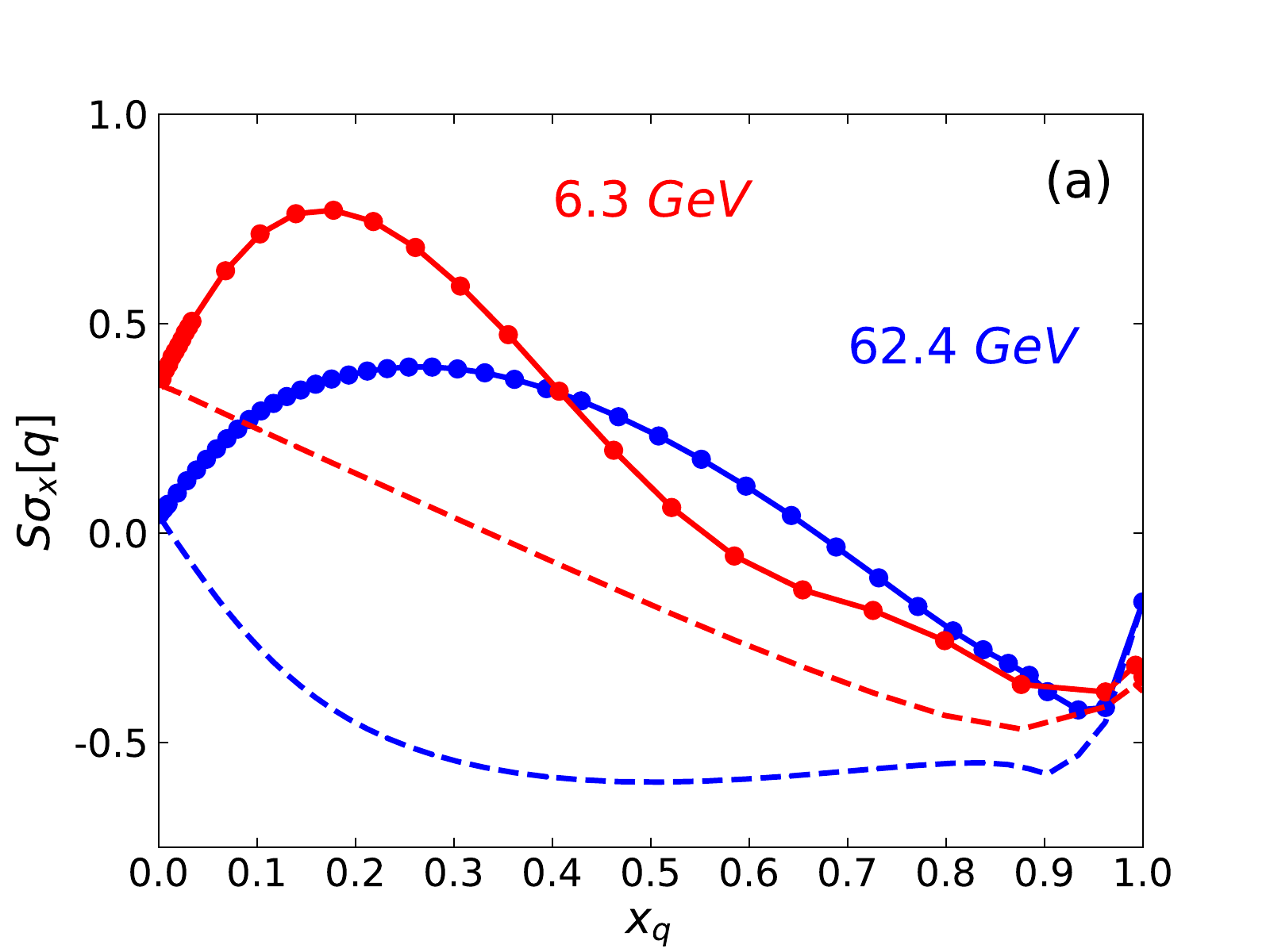}
\includegraphics[width=.49\textwidth]{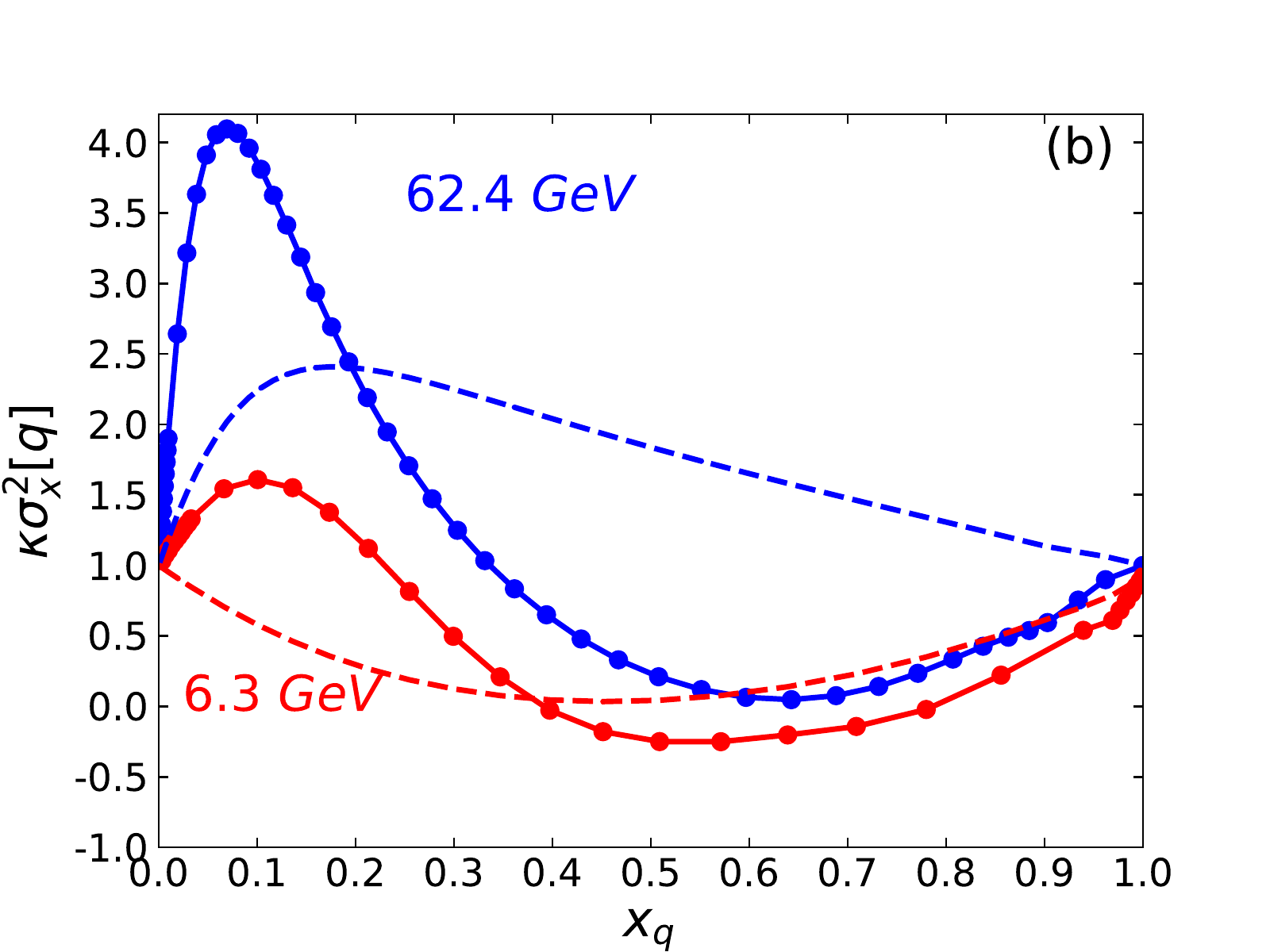}
\includegraphics[width=.49\textwidth]{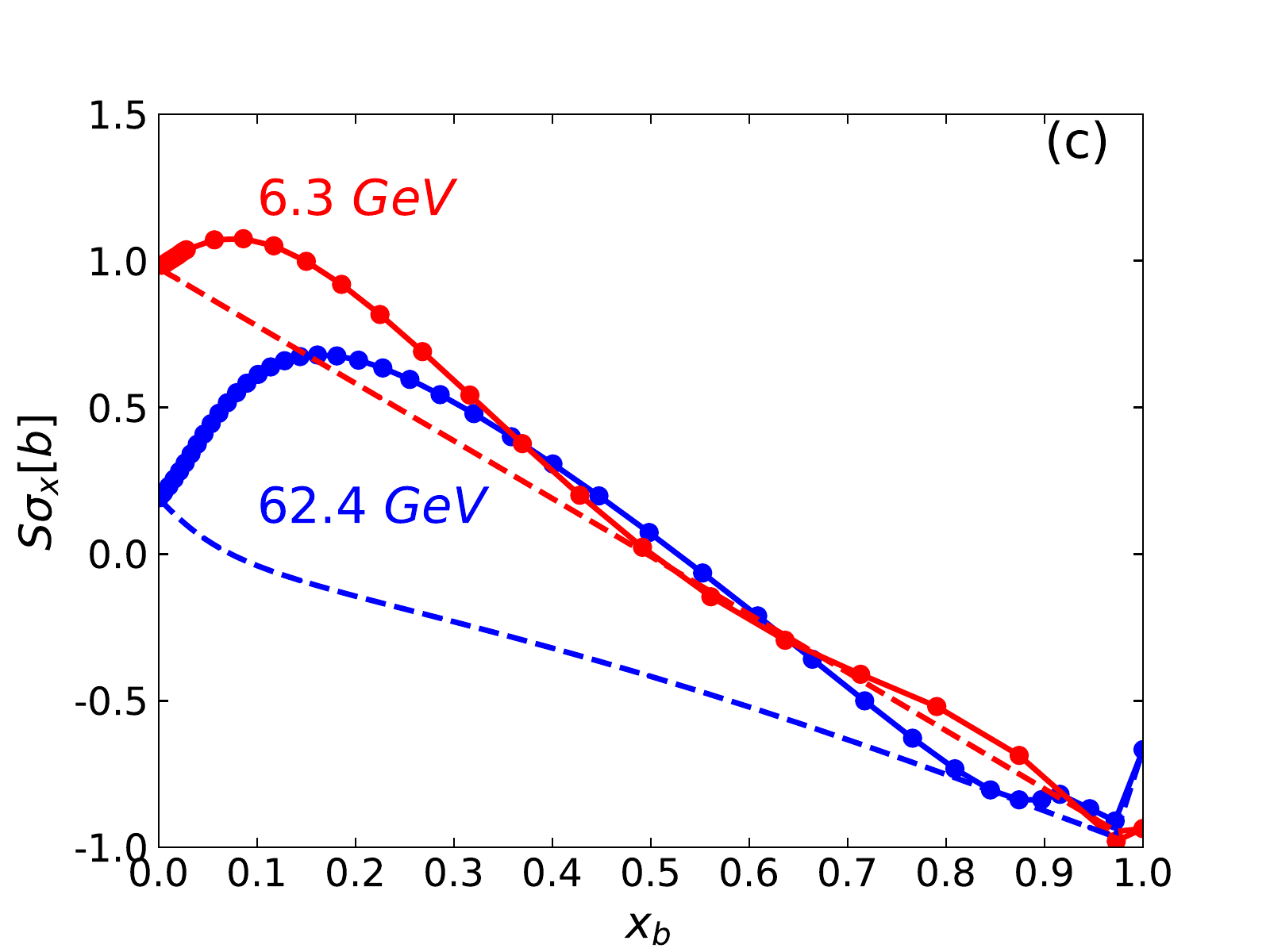}
\includegraphics[width=.49\textwidth]{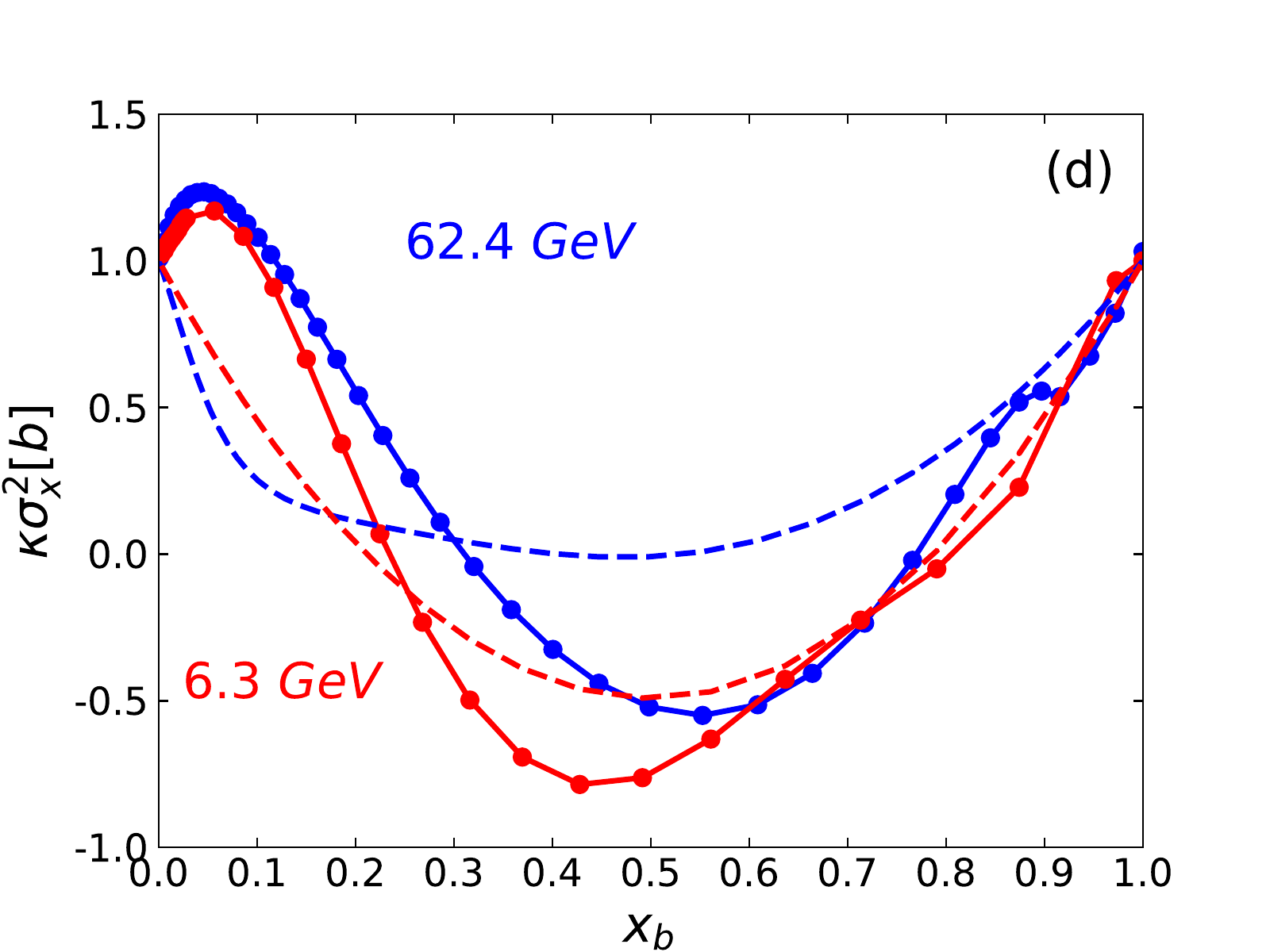}
\caption{\label{fig-charge}
Skewness and kurtosis of the net charge fluctuations ($a$),($b$) and the net baryon number fluctuations ($c$),($d$) as functions of, respectively, net charge acceptance and net baryon number acceptance are calculated in the UrQMD model (blue and red points) and using the BAC (\ref{k1x1x2})-(\ref{k4x1x2}) (dashed lines).
}
\end{figure*}

Figure \ref{vol_versus_rap} (a) and (b)  presents the UrQMD results for binomial acceptances parameters for positively and negatively charged hadrons,  $x_+$ and $x_-$, for baryons and antibaryons, $x_B$ and $x_{\overline{B}}$, as well as for net charge and net baryon numbers
\eq{\label{xq}
x_q \equiv \frac{\langle n_+\rangle-\langle n_-\rangle}{Q}~,~~~~~ x_b\equiv \frac{\langle n_B \rangle-\langle n_{\overline{B}} \rangle}{B} 
}
as functions of $\Delta y$
for p+p collisions at $\sqrt{s}=6.3$~GeV and $\sqrt{s}=62.4$~GeV, respectively.

As seen from Fig.~\ref{vol_versus_rap}, $x_->x_+$ and $x_{\overline{B}}>x_B$. 
This is due to the difference in rapidity spectra, $dN_i/dy$, of the negatively and positively charged hadrons in $p+p$ collisions, and  similar difference of the rapidity spectra of antibaryons and baryons. 
Thus, one cannot apply the simplified BAC equations~(\ref{w_x[Q]}) and (\ref{kq}) which would be valid for $x_+ = x_-$ or $x_B=x_{\overline{B}}$.
The value of $\Delta x\equiv  x_--x_+ $ decreases and goes to zero in the limits $\Delta y\rightarrow 0 $ and $\Delta y \rightarrow \infty$. At large enough $\Delta y$ all particles are accepted. Thus, both $x_-$ and $x_+$ become close to 1, and their difference goes to zero (similar behavior is seen for  $x_{\overline{B}}$ and $x_B$).  
At a low collision energy of $\sqrt{s}=6.3$~GeV there are almost no antibaryons produced, thus $x_b \approx x_{B}$.

\subsection{Scaled variances}
Figure \ref{fig-n-} 
presents the scaled variances for negatively(positively) charged hadrons and for (anti-)baryons accepted in the central rapidity region as functions of the corresponding  acceptance parameters $x_i$. 
The UrQMD results are shown by full blue and red points, and the BAC expressions according to Eqs.~(\ref{w-x})-(\ref{k_x}) are presented  by  dashed lines. 
As expected, the fluctuations measures approach the Poisson limit predicted by the BAC as $x\rightarrow 0$.
Note that this does \emph{not} imply that correlations between particles disappear for small acceptance.
Rather, it is the ability to \emph{measure} these correlations using cumulants which does~(see a related discussion in Ref.~\cite{Pruneau:2019baa}).
The BAC reproduces the UrQMD results in the full phase space limit $x\rightarrow 1$ by construction.
The BAC interpolates $x$-dependence of
$\omega_x[n_i]$  by a straight line, as follows from (\ref{w-x}).
These BAC results are found to deviate considerably from the actual UrQMD results.
For all considered  quantities presented in  Fig.~\ref{fig-n-} the actual UrQMD values for $\omega_x[n_i]$ appear to be larger than those obtained within the BAC procedure. Another interesting feature is an increase of $\omega_x[n_i]$ at small values of $x_i$. This leads to  $\omega_x[n_i]> 1$ 
at small $x_i$, i.e.,  
the fluctuations of $n_i$ at small $x_i$ exceed the Poisson baseline. 
This takes place for all considered particle types, $i=+,-,\overline{B},B$, and for the both collision energies considered. Such a behavior leads to a maximum of $\omega_x[n_i]$ as functions of $x_i$ at intermediate values of $x_i< 1$, and to their decrease at $x_i\rightarrow 1$. 
These features of $\omega_x[n_i]$ seem to be rather general for all UrQMD simulations that we performed. 
A decrease of $\omega[n_i]$ at $x_i\rightarrow 1$ is a natural consequence of the global charge conservation. 
The effect is stronger for $\omega[N_+]$ in comparison to $\omega[N_-]$, and for $\omega[N_B]$ in comparison to $\omega[N_{\overline{B}}]$, reflecting positive net charge and net baryon numbers $Q = B = 2$ in p+p interactions.

\subsection{Skewness and kurtosis of conserved charges fluctuations}
Figure \ref{fig-charge} 
presents the skewness and kurtosis for fluctuations of, respectively, the electric charge, $q\equiv n_+-n_-$, and the net baryon number, $b\equiv n_{B}-n_{\bar{B}}$. These quantities are presented as functions of the corresponding acceptance parameters at the two collision energies, $\sqrt{s}=6.3$ and $62.4$~GeV. 
 
At the ends of the acceptance interval, $x\rightarrow 0$ and $x\rightarrow 1$, 
the BAC Eqs.~(\ref{k1x1x2})-(\ref{k4x1x2}) coincide with the
 UrQMD results. 
The BAC expressions are calculated according to Eqs.~(\ref{k1x1x2})-(\ref{k4x1x2})  and are shown in 
Fig.~\ref{fig-charge} for $0 \leq x \leq 1$
by dashed lines. 
These have a more involved structures than straight lines for (\ref{S[Q]})  for $S\sigma_x[q]$ and  $S\sigma_x[b]$ or symmetric parabolas (\ref{kq}) for $\kappa\sigma_x^2[q]$ and $\kappa\sigma_x^2[b]$, which would both be expected if acceptance parameters were equal for particles and antiparticles. 
Besides, values of the skewness at $x\rightarrow 0$ deviate from the results of Eq.~(\ref{w_x[Q]}).
All these complications of the BAC results in comparison to Eqs.~ (\ref{S[Q]}) and (\ref{kq}) are because of essential corrections from $\Delta x$ terms in Eqs.~(\ref{k1x1x2})-(\ref{k4x1x2}). 

The UrQMD results for $S\sigma_x[q]$ and $S\sigma_x[b]$ presented in 
Fig.~\ref{fig-charge} 
show a non-monotonic dependence on the corresponding acceptance parameters $x_q$ and $x_b$. The behavior of  $\kappa\sigma^2_x[q]$ and $\kappa\sigma^2_x[b]$ is 
even more nontrivial:  they demonstrate a zigzag-like behavior with a maximum at small acceptance parameter $x_{q,b}\cong 0.1$ and a minimum at $x_{q,b}=0.4-0.6$ for both collision energies. 
Such features of the skewness and kurtosis for conserved charges can sometimes lead to their non-monotonic dependencies on the collision energy, even in the absence of any mechanisms for critical fluctuations as is the case for UrQMD.
Note that the results of subsections \ref{sec-urqmd} B and C 
have been obtained using the UrQMD model and are not guaranteed to be unique.
It would therefore be useful to perform the analysis within
other event generators, e.g., PYTHIA,
HIJING, AMPT, and SMASH. 
This can be a subject for future studies.

\subsection{BAC inside a limited phase space}
\label{sec-urqmdv2}

A comparison of the BAC formulas with the actual results of the  UrQMD simulations demonstrate rather large differences. For the fluctuation measures $\omega_x$,
$S \sigma_x$, and $\kappa\sigma^2_x$ calculated in the finite rapidity regions $|y|\le \Delta y / 2$ using the BAC with corresponding acceptance $x$-parameters an agreement with the actual UrQMD results is found at the limits $x\rightarrow 0$ and $x\rightarrow 1$.  
However, at $0<x<1$ the BAC and the actual UrQMD results 
are essentially different, even qualitatively.
This means that certain assumptions behind the BAC procedure are not fulfilled in UrQMD.
The key BAC assumption is that a probability for a given particle to be within acceptance is independent of all other particles.
However, there is an evident reason for event-by-event correlations between the shapes of rapidity distributions and total event multiplicities. From kinematical arguments one expects more final particles just at small center of mass rapidities in events with larger total hadron multiplicities. More special interparticle rapidity correlations emerge in the UrQMD simulations from decays of resonances and strings.
For these reasons, the inaccuracy of the BAC predictions is not all that surprising.

On the other hand, at smaller momentum scales the BAC assumptions might be more reasonable.
Let us take the fixed rapidity interval $\Delta y=2$. 
We will treat now this rapidity interval as a `full phase space' region, and  
the BAC values will be  calculated  
at smaller parts of the rapidity interval $\Delta y=2$. 
The acceptance $x_i$-parameters  is now defined as 
\eq{\label{x-}
x_i = \frac{\langle n_-\rangle_{\Delta y <2}}{\langle n_i\rangle_ {\Delta y=2}}~.
}

\begin{figure*}
\label{BAC_q}
\includegraphics[width=.49\textwidth]{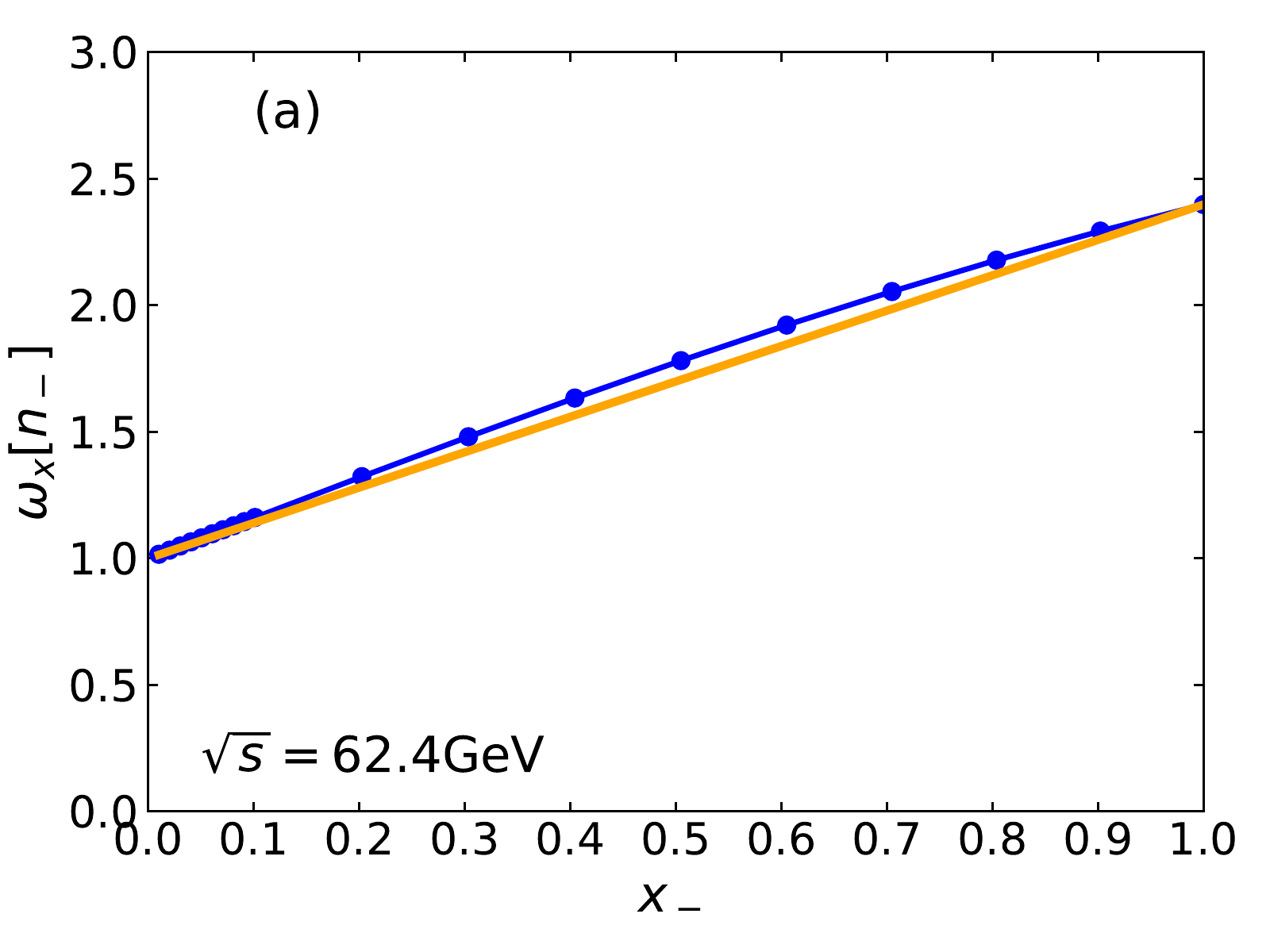}
\includegraphics[width=.49\textwidth]{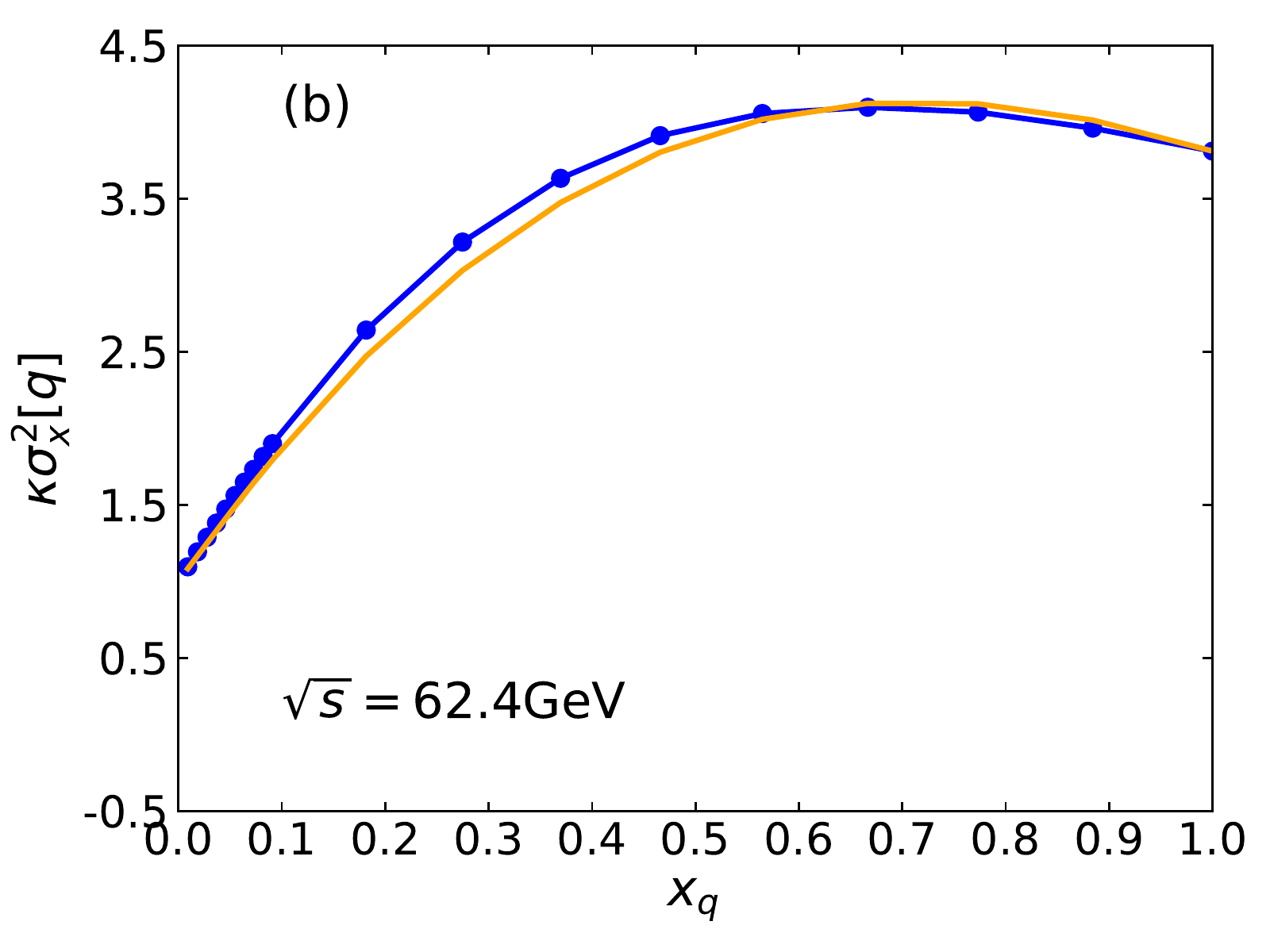}
\caption{\label{fig-test}
$\omega_x[n_-]$ (a) and $\kappa\sigma^2[q]$ (b) inside the rapidity interval $\Delta y<2$ as function of corresponding acceptance parameters. The acceptance parameters  $x_i$ are defined by Eq.~(\ref{x-}). Full blue points correspond to the UrQMD results in $p+p$ reactions at $\sqrt{s}=62.4$~GeV while
orange solid lines present the BAC results.
}
\end{figure*}

As an example, the UrQMD results for $\omega_x[n_-]$ and $\kappa\sigma^2_x[q]$ in $p+p$ reactions at $\sqrt{s}=62.4$~GeV inside the rapidity interval $\Delta y=2$ are shown in 
Fig.~\ref{fig-test}~(a) and (b), respectively. 
As seen from Fig.~\ref{fig-test} (a)  an agreement of the BAC procedure and direct UrQMD results for $\omega_x[n_-]$  inside the rapidity interval $\Delta y=2$ is almost perfect.   
The BAC formula (\ref{w-x}) is used  with  $\langle n_-\rangle_{\Delta y=2}\equiv N_-$ and  $\omega[N_-]=2.4$. 
Similar results were obtained for other fluctuation measures. In Fig.~\ref{fig-test} (b) the same is done for $\kappa\sigma^2_x[q]$ treating all UrQMD quantities at the rapidity interval $\Delta y$ as the 'full phase space' values in the BAC formulas. 
Thus, a basic assumption of the binomial distribution becomes valid for these rapidity intervals and the BAC procedure leads to the results consistent with the actual UrQMD simulations as presented in Fig.~\ref{fig-test}. The local correlations due to the quantum number conservation are present. However, if the experimental acceptance is essentially smaller than the characteristic  length of these correlations, the BAC procedure can provide a good approximation.

It should be noted that the basic question mentioned in Sec.~I -- how to define a suitable acceptance region to observe the {\it statistical} fluctuations of conserved charges within the {\it grand canonical ensemble} -- remains beyond the scope of the present study. It can be wrong to search for the statistical fluctuations in the framework of {\it non-equilibrium} transport model. This is especially clear in $p+p$ reactions at large collision energy. Both the UrQMD results and the $p+p$ reactions data demonstrate large values of particle number fluctuations,  e.g., $\omega[N_{\rm ch}]\propto \langle N_{\rm ch}\rangle  \gg 1$, much above of the standard statistical estimates \cite{Konchakovski:2007ah}.  Applicability of the BAC procedure inside the central rapidity interval, as in Fig.~\ref{fig-test}, is by no means the argument in favor of the statistical character of particle number fluctuations inside this region. In the statistical system treated within the grand canonical ensemble all intensive fluctuation measures remain unchanged in their subsystems, if only these subsystems are not too small compared to the correlation length.  The BAC procedure is fully consistent with such systems only in the simplest case -- a mixture of non-interacting Boltzmann (classical statistics) particles at fixed volume $V$ and temperature $T$. 

\begin{figure*}\label{BAC_b}
\includegraphics[width=.49\textwidth]{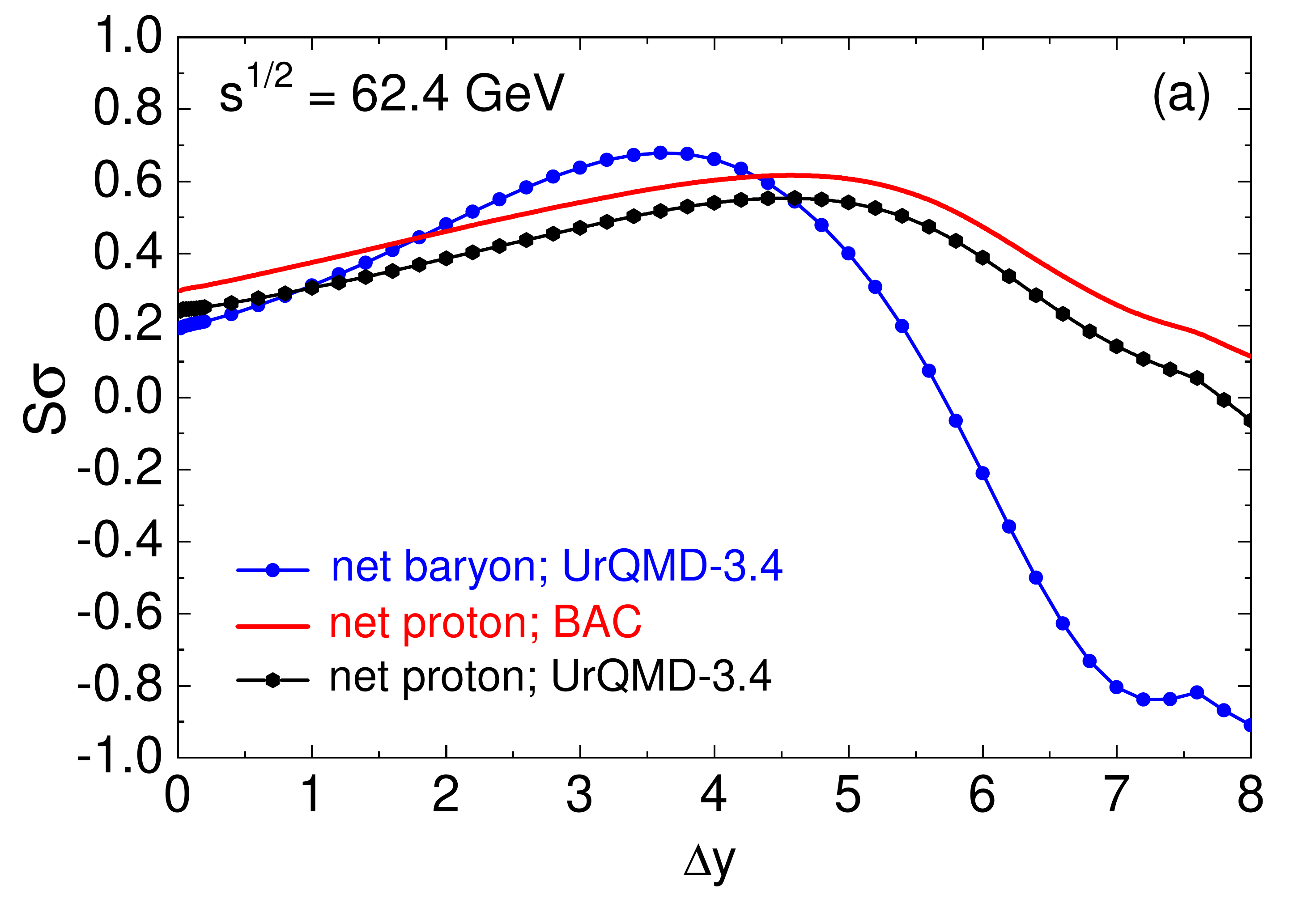}
\includegraphics[width=.49\textwidth]{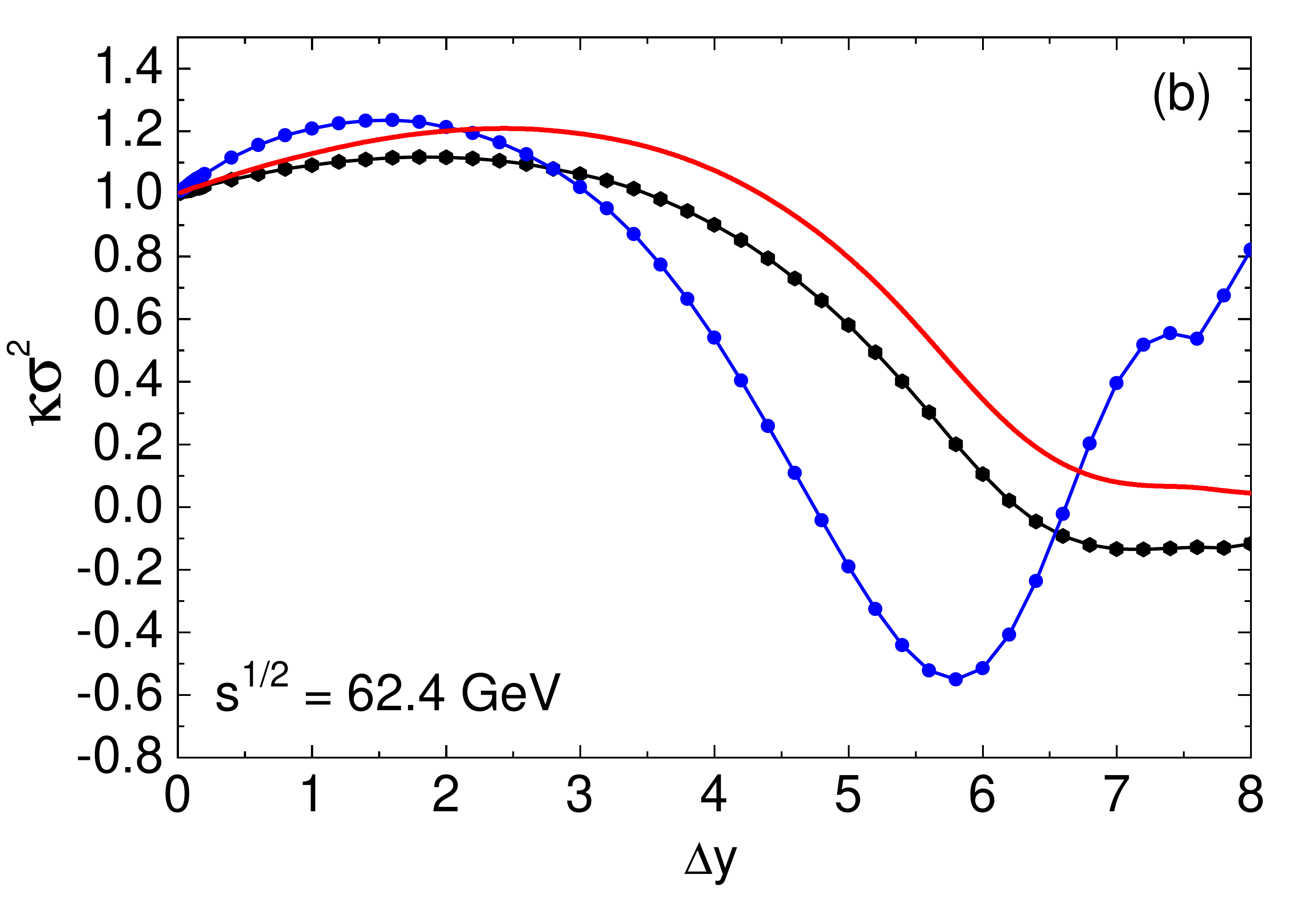}
\caption{\label{fig-efficiency}
Skewness (a) and  kurtosis (b) for inelastic $p+p$ interactions at collision energy of $\sqrt{s}=62.4$~GeV as functions of $\Delta y$. Blue and black solid full points represent, respectively, the net baryon number and the net proton number fluctuations as calculated in the UrQMD model. Red solid 
lines represent the net proton number fluctuations as obtained using the BAC
formulas with $x_p$ and $x_{\overline{p}}$  given by Eq.~(\ref{x-p}) and
fixed values of $x_{p/B}=0.5$, $x_{\bar{p}/\bar{B}}=0.4$.
}
\end{figure*}

\subsection{Net proton fluctuations from net baryon fluctuations}

One interesting question
is a connection of (anti-)proton fluctuations with those of (anti-)baryons. 
In the experiment, the skewness and kurtosis of the net proton number fluctuations are measured, but not of the net baryon number ones due to the problems with detecting neutral baryons and anti-baryons, mainly neutrons and anti-neutrons. 
One can consider now BAC by assuming that a randomly chosen baryon is within an acceptance if it is a proton and outside of it otherwise.
In $p+p$ reactions at $\sqrt{s}=62.4$~GeV the UrQMD results in the full phase space provide:
\eq{\label{x-p}
x_p\equiv \frac{\langle N_p\rangle }{\langle N_B\rangle}~\approx 0.5, ~~~~~x_{\overline{p}}\equiv \frac{\langle N_{\overline{p}}\rangle }{\langle N_{\overline{B}}\rangle}~\approx 0.4~,
}
where $N_p$ and $N_{\overline{p}}$ denote the numbers of protons and antiprotons, respectively.

Figure~\ref{fig-efficiency} presents the skewness and kurtosis of the net baryon and the net proton number fluctuations in $p+p$ reactions at $\sqrt{s}=62.4$~GeV.
The UrQMD results, depicted by lines with symbols, show sizable differences between net baryon and net proton fluctuation measures, suggesting that the latter may not be a particularly good direct proxy for the former.
The red line shows the net proton fluctuations constructed out of the net baryon fluctuations by applying the BAC using the $x_p$ and $x_{\bar{p}}$ acceptance parameters listed above.
One observes a quite good agreement of the net-baryon fluctuations calculated by the BAC procedure from the net-baryon values using acceptance parameters (\ref{x-p}) with their exact UrQMD values. 
The results suggest that constructing  the net proton fluctuations from net baryon ones using a binomial filter, as suggested in Refs.~\cite{Kitazawa:2011wh,Kitazawa:2012at}, might be a reasonable procedure. This observation is important in the context of attempts to relate the net proton fluctuation measurements in heavy-ion collisions with QCD net baryon number susceptibilities computed e.g. using first-principle lattice simulations~\cite{Bazavov:2017tot}.
 We hope that the results of subsections \ref{sec-urqmd} D and E is not due to the specific features of the UrQMD model and can be also approximately valid within the data. In order to check this, analogous calculations can be performed using other models, which can be a subject for future studies.

\section{summary}
\label{summary}

We studied the binomial acceptance corrections which relate distributions of various particle number and/or conserved charge distributions in a region of phase space to the corresponding distributions in the full phase space.
The binomial acceptance corrections are derived under the assumption that each particle is accepted with a certain probability independently from all other particles.
Based on this, we derive explicit formulas that connect high order cumulants and their various ratios such as scaled variance, skewness and kurtosis of fluctuations within a given acceptance~(sub-system) with those in a broader acceptance~(system).
Where applicable, our formalism reproduces earlier results on the binomial acceptance~\cite{Kitazawa:2011wh,Bzdak:2012ab}.

The BAC transform a Poisson distribution into another Poisson distribution. Therefore BAC cancels out in all cumulant ratios if the underlying particle number distribution in the full space is Poissonian.
However, this is not the case for other distributions.
Particularly the fluctuations of conserved charges, i.e. quantities which are conserved globally, are studied in some detail in the present work.
Exact conservation induces correlations between positive and negative particles, in contrast to the Poisson baseline that entails no correlations.
We show that fluctuations of these quantities within a given acceptance are expressed through fluctuations of a sum of positively and negatively charged particles $N_{\rm ch} = N_+ + N_-$ in the full phase space, Eqs.~\eqref{k1x1x2}-\eqref{k4x1x2}.
As a particular example, we explore $N_{\rm ch}$ and $N_{\pm}$ distributions within canonical relativistic statistical mechanics. In contrast to the grand-canonical ensemble where all these fluctuations are the Poisson ones, in the canonical ensemble, these are given by a more involved Bessel distribution.
These observations will be useful for future measurements and analysis of net electric charge fluctuations.
It should be noted, however, that a presence of local charge conservation and radial flow can break the assumptions behind the binomial acceptance procedure, namely that the momenta of particles are uncorrelated.
Such questions might be better clarified by measurements of momentum dependent balance functions, as suggested recently in Ref.~\cite{Pruneau:2019baa}.

UrQMD simulations of inelastic p+p interactions were then used to explore the performance of the BAC when applied to a momentum space acceptance, as is appropriate for high-energy collision experiments.
It was found that actual UrQMD fluctuations of various particle numbers in a given rapidity window deviate considerably from those predicted using the BAC procedure applied to fluctuations in full phase space. 
This indicates that the BAC assumption of an uncorrelated acceptance probability is not fulfilled in UrQMD simulations, which can take place if particle rapidities are correlated on scales smaller than beam rapidities.
We do find that the BAC procedure is found to be significantly more accurate when applied to relate fluctuations between various smaller rapidity windows which span no more than two units.
We plan to address system-size systematics of the BAC accuracy in a future work.

The BAC can also be used 
to correct for
the inability to measure cumulants of neutral particles such as neutrons.
Our UrQMD analysis of p+p collisions shows that net proton fluctuations obtained by applying the BAC to net baryon fluctuations agrees quite well with actual net proton fluctuations.
The BAC can thus be used to reconstruct the net baryon fluctuations from the measured net proton ones~\cite{Kitazawa:2011wh,Kitazawa:2012at} or to estimate the net proton fluctuations in a framework where their explicit calculation is problematic but where baryon and antibaryon fluctuations are tractable by theoretical models.

\section*{Acknowledgments}
The authors thank  Marek Gazdzicki,  Volker Koch, Maja Maćkowiak-Pawłowska, Anton Motornenko, Anar Rustamov, and Horst Stoecker for fruitful discussions and useful comments.
The work of M.I.G. is supported by the Department of Physics and Astronomy of the National Academy of Sciences of Ukraine.

\end{widetext}


%

\end{document}